\documentclass{emulateapj}

\shorttitle{WMAP extragalactic point sources}
\shortauthors{L\'opez-Caniego et al.}

\begin{document}

\title{Non-blind catalogue of extragalactic point sources from \\
the Wilkinson Microwave Anisotropy Probe (WMAP) \\
first 3--year survey data}

\author{M. L\'opez-Caniego}
\affil{Instituto de F\'isica de Cantabria (CSIC-UC), Santander, Spain 39005}
\affil{Departamento de F\'isica Moderna, Universidad de Cantabria, Santander, Spain 39005}
\email{caniego@ifca.unican.es}

\author{J. Gonz\'alez-Nuevo} 
\affil{SISSA-I.S.A.S, Trieste, Italy 34014}

\author{D. Herranz} 
\affil{Instituto de F\'isica de Cantabria (CSIC-UC), Santander, Spain 39005}

\author{M. Massardi} 
\affil{SISSA-I.S.A.S, Trieste, Italy 34014}

\author{J. Luis Sanz} 
\affil{Instituto de F\'isica de Cantabria (CSIC-UC), Santander, Spain 39005}

\author{G. De Zotti} 
\affil{INAF-Ossevatorio Astronomico di Padova, Padova, Italy 35122}

\author{L. Toffolatti} 
\affil{Departamento de F\'isica, Universidad de Oviedo, Oviedo 33007}

\and

\author{F. Arg\"ueso}  
\affil{Departamento de Matem\'aticas, Universidad de Oviedo, Oviedo 33007}

\begin{abstract}
We have used the MHW2 filter to obtain estimates of the flux densities
at the WMAP frequencies of a complete sample of 2491 sources, mostly
brighter than 500 mJy at 5 GHz, distributed over the whole sky
excluding a strip around the Galactic equator ($|b|\le
5^\circ$). After having detected 933 sources at the $\geq 3\sigma$
level in the MHW2 filtered maps - our New Extragalactic WMAP Point
Source (NEWPS)$_{3\sigma}$ Catalogue - we are left with 381 sources at
$\geq 5\sigma$ in at least one WMAP channel, 369 of which constitute
our NEWPS$_{5\sigma}$ catalogue. It is remarkable to note that 98
(i.e. 26\%) sources detected at $\geq 5\sigma$ are `new'', they are
not present in the WMAP catalogue. Source fluxes have been corrected
for the Eddington bias. Our flux density estimates before such
correction are generally in good agreement with the WMAP ones at 23
GHz.  At higher frequencies WMAP fluxes tend to be slightly higher
than ours, probably because WMAP estimates neglect the deviations of
the point spread function from a Gaussian shape. On the whole, above
the estimated completeness limit of 1.1 Jy at 23 GHz we detected 43
sources missed by the blind method adopted by the WMAP team. On the
other hand, our low-frequency selection threshold left out 25 WMAP
sources, only 12 of which, however, are $\ge 5\sigma$ detections and
only 3 have $S_{23\rm GHz} \ge 1.1\,$Jy. Thus, our approach proved to
be competitive with, and complementary to the WMAP one.
\end{abstract}


\section{Introduction} \label{sec:intro}

As a by-product of its temperature and polarization maps of the Cosmic
Microwave Background (CMB), the Wilkinson Microwave Anisotropy Probe
(WMAP) mission has produced the first all-sky surveys of extragalactic
sources at 23, 33, 41, 61 and 94 GHz (Bennett et al. 2003; Hinshaw et
al. 2006), a still unexplored frequency region where many interesting
astrophysical phenomena are expected to show up (see e.g. De Zotti et
al. 2005).

>From the analysis of the first three years survey data the WMAP team
has obtained a catalogue of 323 extragalactic point sources (EPS;
Hinshaw et al. 2006), substantially enlarging the first-year one that
included 208 EPS detected above a flux limit of $\sim 0.8$-1 Jy
\citep{ben03b}.

As discussed by \citet{hinshaw06}, the detection process -- quite
similar to the one adopted for producing the first-year catalogue
-- can be summarized as follows. First, the weighted map,
$N_{obs}^{1/2}T$, where $N_{obs}$ is the number of observations
per pixel, is filtered in the harmonic space by the global matched
filter,
\begin{equation} \label{eq:global_MF}
  \frac{b_{\ell}}{b^2_{\ell}C_{\ell}^{\rm cmb}+C_{\ell}^{\rm noise}},
\end{equation}
\noindent where $b_{\ell}$ is the transfer function of the WMAP
beam response \citep{page03,jarosik06}, $C_{\ell}^{\rm cmb}$ is
the angular power spectrum of the CMB and $C_{\ell}^{\rm noise}$
represents the noise power. Then, all the peaks above the
$5\sigma$ threshold in the filtered maps are assumed to be source
detections and are fitted in real space, i.e. in the unfiltered
maps, to a Gaussian profile plus a planar baseline. The Gaussian
amplitude is finally converted to a flux density using the
conversion factors given in Table 5 of \citet{page03}. The error
on the flux density is given by the statistical error of the fit.

If a source has been detected above $5\sigma$, $\sigma$ being the
noise rms defined globally, in a frequency channel, the flux
densities in the other channels are given if they are above
$2\sigma$ and the source width falls within a factor of two of the
\emph{true} beam width. In other words, not all sources listed at
any given frequency in the three-year WMAP catalogue of EPS
\citep[Table 9]{hinshaw06} are $5\sigma$ detections at that
frequency. Finally, to identify the detected sources with 5 GHz
counterparts, they cross-correlated them with the GB6, PMN and
\citet{kuehr81} catalogues.

The fact that almost all sources detected by WMAP were previously
catalogued at lower frequencies suggests that a fuller
exploitation of the WMAP data can be achieved complementing the
\emph{blind} search already carried out by the WMAP team with a
search for WMAP counterparts to sources detected at lower
frequencies. The latter approach exploits the knowledge of source
positions to extract as much information as possible on their
fluxes.

Since point sources, as any other foreground emission, are a
nuisance for CMB experiments, these are designed to keep them as
low as possible. Thus, if we are interested in point sources, it
is of great importance to develop and apply specific and highly
efficient detection algorithms. It has been shown that wavelet
techniques perform very well this task
\citep{cay00,patri01a,patri01b,patri03,jgn06,can06}.

In a recent paper \citep{jgn06} some of us have discussed a natural
generalization of the (circular) Mexican Hat wavelet on $R^2$,
obtained by iteratively applying the Laplacian operator to the
Gaussian function, which we called the Mexican Hat Wavelet Family
(MHWF). We demonstrated that the MHWF performs better than the
standard Mexican Hat Wavelet \citep[MHW,][]{cay00} for the detection
of EPS in CMB anisotropy maps.  In a subsequent work \citep{can06} the
MHWF has been applied to the detection of compact extragalactic
sources in simulated CMB maps.  In that work it was shown, in
particular, that the second member of the MHWF, called the MHW2, at
its optimal scale compares very well with the standard Matched Filter
(MF); its performances are very similar to those of the MF and it is
much easier to implement and use\footnote{\citet{can06} compared the
two techniques, MF vs.  MHW2, by exploiting realistic simulations of
CMB anisotropies and of the Galactic and extragalactic foregrounds at
the nine frequencies, between 30 and 857 GHz, of the ESA Planck
mission, considering the goal performances of the Planck Low and High
Frequency Instruments, LFI and HFI \citep[see][]{pbbook}.}. We have
therefore decided to apply the MHW2 technique to estimate the flux of
the EPS in the 3-year WMAP maps for a complete sample of sources
selected at lower frequencies.

The outline of the paper is as follows. In Section~\ref{sec:5ghz},
we introduce the low-frequency selected sample we have used.  In
Section~\ref{sec:method} we present and discuss the tools for
detecting the EPS and for estimating their fluxes.  In
Section~\ref{sec:sims} we present and describe our final
catalogue. In Section~\ref{sec:results} we compare our catalogue
with the WMAP one. Finally, in Section~\ref{sec:Conclusions}, we
summarize our main conclusions.

\section{The 5 GHz catalogue} \label{sec:5ghz}

\begin{figure}
\begin{center}
\includegraphics[width=0.45\textwidth]{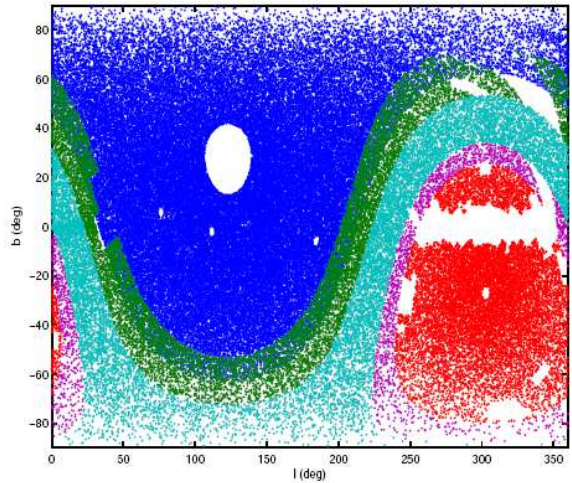}
\caption{Sky coverage of 5 GHz surveys in equatorial coordinates:
GB6 \citep{gre96} (blue), PMNE \citep{gri95} (dark green), PMNS
\citep{wri94} (red), PMNT \citep{gri94} (light blue) and PMNZ
\citep{wri96} (magenta). The white regions are `holes' in these
surveys, that have been covered exploiting the NVSS and the SUMSS.
\label{sky_cov}}
\end{center}
\end{figure}

A summary of the multi--steradian surveys that we have used is given
in Table~\ref{tb:summary}. The highest frequency for which an almost
complete sky coverage has been achieved is $\simeq 5\,$GHz, thanks to
the combined 4.85 GHz GB6 and PMN surveys with an angular resolution
of $3.5^{\prime}$ and $4.2^{\prime}$, respectively, and a flux limit
ranging from 18 to 72 mJy. The sky coverage of these surveys is
illustrated in Fig.~\ref{sky_cov}.  Deeper and higher resolution
surveys have been carried out at 1.4 (NVSS, Condon et al. 1998; FIRST,
Becker et al. 1995) and 0.843 GHz (SUMSS, Mauch et al. 2003);
altogether these surveys cover the full sky.

As extensively discussed by many authors in the recent past
\citep{tof99,dz99,mas03,ben03b,h&p05}, ``flat--spectrum'' AGNs and
QSOs, i.e. sources showing a spectral index $\alpha\simeq 0$
($S(\nu )\propto \nu^{-\alpha}$), are expected to be the dominant
source population in the range 30--100 GHz, whereas other classes
of sources, and in particular the steep-spectrum sources
increasingly dominating with decreasing frequency, are only giving
minor contributions to the number counts at WMAP frequencies and
sensitivities \citep{dz06}. We therefore chose to adopt 5 GHz as
our reference frequency, and used lower frequency surveys to fill
the ``holes'' at 5 GHz.

\begin{figure*}
  \begin{center}
    \includegraphics[width=10cm]{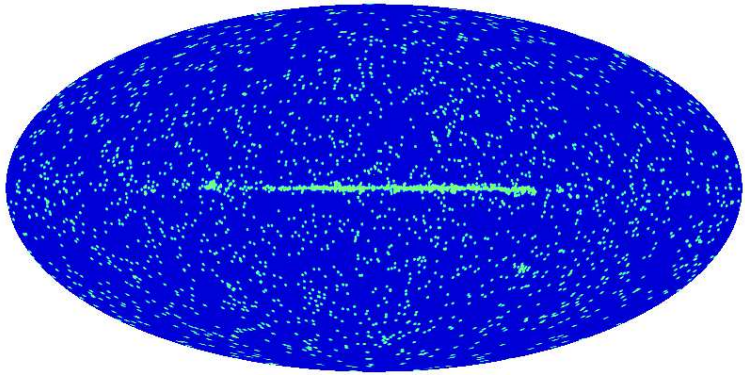}
    \caption{Sky distribution, in Galactic coordinates, of the 4050
      objects with flux densities above our selection threshold. \label{fig:IC}}
  \end{center}
\end{figure*}

Altogether, the catalogues listed in Table~\ref{tb:summary} contain
over 2 million sources, but we already know, from the analysis of the
WMAP team, that for only a tiny fraction ($\sim 2\times 10^{-4}$) of
them the WMAP data can provide useful information. Applying the MHW2
at the positions of all these sources would not only be extremely
inefficient, but plainly unpractical because of the huge CPU time and
disk storage requirements.  Therefore, we decided to work with a
complete sub-sample containing sources with $S_{5 \rm GHz} \ge
500\,$mJy.  This limiting flux corresponds to about 2--3 times the
mean noise in the filtered images we will be dealing with (see
\S\,\ref{sec:method}). To fill the 5 GHz ``holes'' we have picked up
NVSS or SUMSS (in the region not covered by the NVSS) sources brighter
than 500 mJy at the survey frequency. In this way we obtained an
all-sky sub-sample of 4050 objects, whose spatial distribution, in
Galactic coordinates, is shown in Fig.~\ref{fig:IC}.  After having
removed sources in the strip $|b|\le 5^\circ$, and in the LMC region
(i.e. inside the circle of $5.5^\circ$ radius centered at $\alpha= 5^h
23^m 34^s.7$, $\delta = -69^\circ 45' 22''$, J2000; $l =
280.47^\circ$, $b = -32.89^\circ$) and the Galactic sources outside of
these zones (Taurus A, Orion A \& B, and the planetary
nebula IC\,418/PMNJ0527-1241) we are left with 2491 objects making up
our ``Input Catalogue'' (IC).

A cross-correlation of the IC with the WMAP catalogue
\citep{hinshaw06} with a search radius of $20.8'$, equal to the
dispersion of the Gaussian approximation of the beam of the lowest
resolution WMAP channel ($23$ GHz), showed that $298$ of the $323$
WMAP sources have a counterpart in the IC. The other 25 WMAP sources
(called \emph{missed sources}) must be unusually weak at low frequency,
either because have an inverted spectrum or are strongly variable and
were caught in a bright phase by WMAP. They are thus interesting targets
for further study, and we have investigated them too. As they have a
different selection, these sources are listed separately from the
others (Table 4).

\section{Methodology}   \label{sec:method}

The strategy used by the WMAP team to obtain the catalogue of 323
point sources \citep{hinshaw06}, summarized in
\S$\,$\ref{sec:intro}, used some approximations:
\begin{itemize}
\item[-] The candidate detections were selected as $5\sigma$ peaks,
where $\sigma$ is the rms noise defined globally. However, the
removal of the Galactic emissions by component separation methods
is not perfect, and leaves non-uniform contributions to the noise.
Clearly, the detection, identification and flux estimation of EPS
can benefit from a local treatment of the background. Furthermore,
considering a global rms noise can lead to an underestimate of the
error in regions of strong Galactic emission.
\item[-] The intensity peaks were fitted to a Gaussian profile.
However, if the real beam response function is non-Gaussian,
   a Gaussian fit can lead to systematic errors in the flux estimate.
\item[-] Confusion due to other point sources that are close to
  the target one, albeit a less relevant effect, is another source of error
  that makes necessary to study the data locally.
\end{itemize}
As mentioned above, we used, locally, the MHW2 that proved to be
as efficient as the matched filter but easier to implement and
more stable against local power spectrum fluctuations
(Gonz\'alez-Nuevo et al. 2006; L\'opez-Caniego et al. 2006). The
real symmetrized radial beam profiles given by the WMAP team have
been used rather than their Gaussian approximations. The error on
the source flux density is calculated {\it locally} as the rms
fluctuations around the source. Finally, we also corrected the
flux of all the sources for the Eddington bias, adopting a
Bayesian approach (see \S\,\ref{sec:bayes}).

\subsection{From the sphere to flat patches: rotation \&
projection}\label{sec:rot}

In order to avoid CPU and memory expensive iterative filtering in
harmonic space we chose to work with small flat sky patches. For every
source position in the IC and for each WMAP frequency we obtained a
flat patch of approximately $14.6^\circ \times 14.6^\circ$ size, by
projecting the WMAP full-sky maps. The adopted pixel area is $6.87
\times 6.87 \,\hbox{arcmin}^2$ (NSIDE=512), so that the patches are made of $128
\times 128$ pixels. The patch making goes as follows:
\begin{itemize}
\item[-] Given the source coordinates we obtain the
  corresponding pixel in the HEALPix \citep{gorski05} scheme.
\item[-] The image is rotated in the $a_{\ell m}$ space so that the
  position of the point source is moved to the equatorial plane. This
  is done in order to minimize the distortion induced by the projection
  of the HEALPix non-square pixels into flat square pixels.
\item[-] The pixels in the sphere in the vicinity of the
  centre are projected using the flat patch approximation and reconstructed into a
  2D image in the plane.
\item[-] The units of the images are converted from mK (WMAP units)
to Jy/sr and finally to Jy, using the real symmetrized radial beam
profiles to do the integrals over the beam area.
\end{itemize}
The rotation/projection scheme described above introduces some
distortions in the projected image. There are two different effects to
be considered: a) the distortion introduced by the projection (from
HEALPix pixels to flat pixels in the tangent plane); b) the distortion
caused by the rotation (from HEALPix pixels to $a_{\ell m}$ and then
to HEALPix pixels again). We have studied these distortions by
simulating 840 sources having flux densities ranging from 500 mJy to
20 Jy with the real WMAP beam profile. First they are simulated
without noise and then they are added on the combined WMAP maps,
placing them at different Galactic longitudes and latitudes. Applying
to each of the simulated sources the same rotation/projection
procedure as for real sources we have found that, as expected, the
projection effects are small at the image center (near the tangent
point) and grow towards the borders of the patch. Also distortion
effects are small near the equator (where the HEALPix pixels are very
close to squares) and grow towards the poles. This is the main reason
for performing the rotation in order to always have the point source
at the equator. We find that the rotation in the $a_{\ell m}$ space
from the initial position of the source to the centre of the map
($l$=0, $b$=0) has a very small effect on the flux estimate (always $<
2\%$), while the effect of the projection is totally negligible.

\subsection{Filters}

\begin{figure*}
  \centering
  \includegraphics[width=0.45\textwidth]{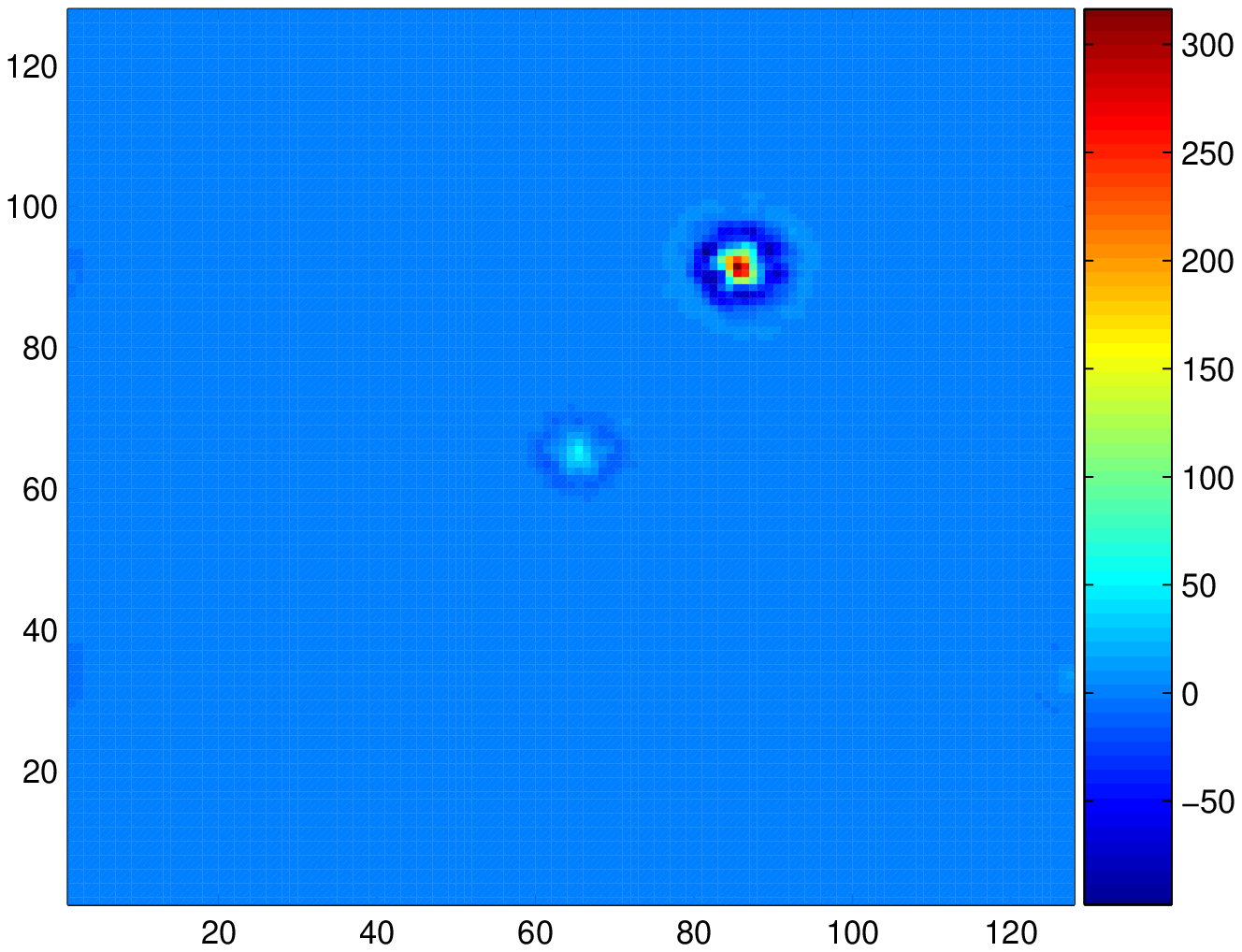}%
  \hfill\includegraphics[width=0.45\textwidth]{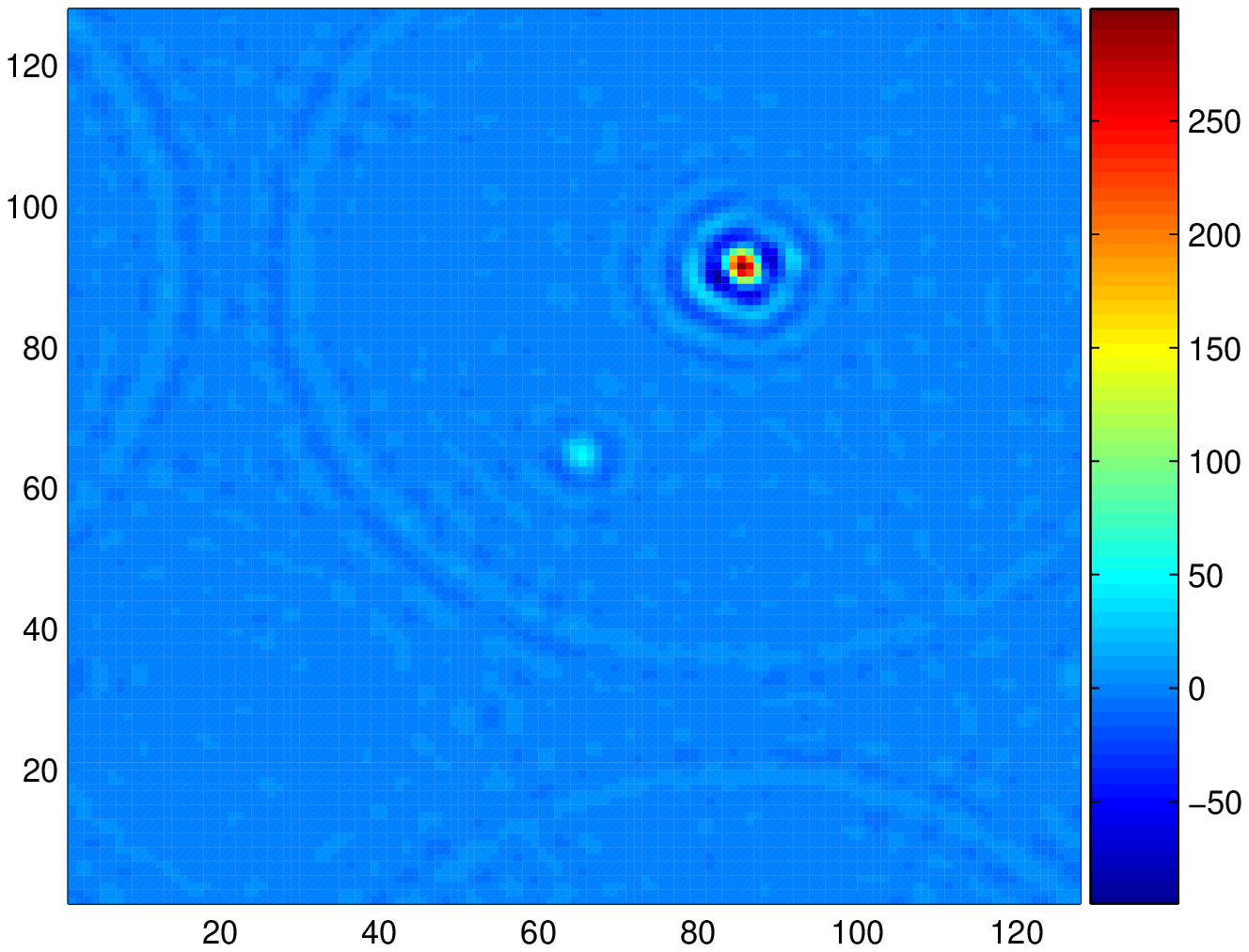}
  \caption{Example of a bright source filtered with the MHW2
  (\emph{left panel}) and the MF (\emph{right panel}).  Strong ringing
  effects can be seen in the case of the MF.}
\label{ringing}
\end{figure*}

In previous works
\citep{sanz99,cay00,sanz01,patri01a,patri01b,patri03,yo02a,yo02b,yo02c,jgn06,can06}
we have shown that application of an appropriate filter to an image
helps a lot in removing large scale variations as well as most of the
noise. As a result, the signal--to--noise ratio is increased, thus
amplifying the point source signal. A brief summary of the
fundamentals of linear filtering of two-dimensional (2D) images is
given in Appendix~\ref{app:filters}. In this work we have only taken
into account two possible filters: the Matched Filter (MF) and the
second member of the Mexican Hat Wavelet Family (MHW2). By applying
these two filters to WMAP temperature maps we find that both of them
give an average amplification of the EPS flux by a factor of $\sim
3$. This is a remarkable result. It means that the flux of a point
source at the $\sim 2\sigma$ level in the original map can be enhanced
to a $\geq 5\sigma$ level in the filtered map. In the following
sub-sections, we briefly sketch the main properties of the MF and of
the MHW2.

\subsubsection{The Matched Filter (MF)}

The MF is a circularly-symmetric filter, $\Psi(x;R,b)$, such that
the filtered map, $w(R, b)$ (see Appendix A for the definition of
the notation), satisfies the following two conditions: $(1) \
\langle w(R_0, 0)\rangle = s(0) \equiv A$, i.e. $w(R, 0)$ is an
\emph{unbiased} estimator of the flux density of the source; $(2)$
the variance of $w(R, b)$ has a minimum on the scale $R_0$, i.e.
it is an \emph{efficient} estimator. In Fourier space the MF
writes:
\begin{equation} \label{eq:mf}
  \psi_{MF} = \frac{1}{a}\frac{\tau (q)}{P(q)}, \ \ \
  a = 2 \pi \int dq q \frac{\tau^2 (q)}{P(q)},
\end{equation}
\noindent where $P(q)$ is the power spectrum of the background and
$\tau (q)$ is Fourier transform of the source profile (equal to
the beam profile for point sources). The matched filter gives
directly the maximum amplification of the source and it yields the
best linear estimation of the flux, when used properly and under
controlled conditions. As mentioned in \citet{can06}, the
practical implementation of the MF requires the estimation of the
power spectrum $P(q)$ directly from the data and this leads to a
certain degradation of its performance.

The WMAP team have done a \emph{global} implementation of the MF on
the sphere by taking into account the {non-Gaussian} profile of the
beam (although the source fluxes are estimated fitting the source
profiles with a Gaussian). The resulting matched filter is given by
eq.~(\ref{eq:global_MF}), where the flat limit quantities $\tau(q)$,
$P(q)$ are replaced by their harmonic equivalents $b_{\ell}$,
$C_{\ell}$.  The use of the $C_\ell$ of the whole sky to construct a
MF filter that operates in the sphere is a good first approach to
obtain a list of source candidates and to estimate their fluxes, but
we do believe that it can be improved by operating locally.  In this
paper we use different filters for regions with different levels of
Galactic contamination.

\subsubsection{The Mexican Hat Wavelet Family (MHWF): MHW2}

One example of wavelet that is particularly well suited for point
source detection is the MHW2, a member of the MHWF first
introduced by \citet{jgn06}. The MHW2 is obtained by applying
twice the Laplacian operator to the Gaussian function. This
wavelet filter operates locally and is capable of removing
simultaneously the large scale variations introduced by the
Galactic foregrounds as well as the small scale structure of the
noise. Note that the expression of the filter can be obtained
analytically (while the MF depends on the $P(q)$ that must be
estimated numerically \emph{and} locally\footnote{Some controversy
has arisen lately on whether the dependence of the MF upon $P(q)$
is a problem from the practical point of view, or not. The main
argument supporting the opinion that there is no problem at all
goes as follows: the angular power spectrum of the background
signal is determined by the CMB, whose power spectrum is fairly
well known, and by the instrumental noise, whose statistics is
perfectly known. This is not quite true since the background
signal includes Galactic emission (or its residual after component
separation), which shows strong variations from one point of the
sky to another, and unresolved point sources. Thus the angular
power spectrum is \emph{not} perfectly known, at least locally. It
must be estimated from the data with some error \emph{that will
inevitably propagate to the filter}. For a more detailed
discussion, see \citet{can06}.}). Any member of this family can be
described in Fourier space as
\begin{equation}
  \psi_{n}(q) \propto q^{2n}e^{-q^{2}/2}.
\end{equation}
The expression in real space for these wavelets is
\begin{equation}
  \label{cc1} \psi_{n}(x) \propto \triangle^{n}{\varphi(x)},
\end{equation}
\noindent where $\varphi$ is the 2D Gaussian
$\varphi(x)=(1/2\pi)e^{-x^2/2}$.  We remark here that the first
member of the family, $\psi_{1}(x)$, is the usual Mexican Hat
Wavelet (MHW), that has been exploited for point source detection
with excellent results \citep{cay00,patri01a,patri03}. Note that
we call MHW$n$ the member of the MHWF with index $n$.

As already mentioned, in this work we filter our projected WMAP
sky patches with the second member of the family, the MHW2. As in
\citet{can06}, we will do a qualitative comparison with the
results obtained with the MF, implemented to be used locally in
flat patches (at variance with the \emph{global} MF used by the
WMAP team).

\subsection{Position, flux and error estimation} \label{sec:pfe}

We want to obtain an estimate of the flux density, with its error, of
the IC sources at the center of each filtered image. Point sources
appear in the image with a profile identical to the beam profile. For
example, if the beams were Gaussian, the ratio between the beam and
the pixel area, $2\pi(R_s)^2/L_p^2$, where
$R_s=\mathrm{FWHM}/(2\sqrt{2\log{2}})$ is the beam width and $L_p$ is
the pixel side, would allow us to convert the flux in the pixel where
the source is located into the source flux.

In our case, the beams are not Gaussian and we will calculate this
relationship integrating over the real beam profile for each
channel. In carrying out the calculation we have to take into
account that we work with HEALPix coordinates, at the WMAP
resolution. Although the image is centered on the source position,
after the projection to the flat patch the source does not always
lay in the central pixel, but may end up in an adjacent one. Thus,
to estimate its flux we make reference not to the intensity in the
central pixel but to that of the brightest pixel close to the
center of the \emph{filtered} image (however in most cases the
brightest pixel coincides with the central one).

At first glance, this may seem a very crude estimator, but it
turns out that flux estimation through linear filtering is almost
optimal in many circumstances. Let us explain how this estimator
works. After filtering, the intensity of the brightest pixel can
be written as a weighted sum of the intensities in the surrounding
pixels,

\begin{equation}\label{eq:conv_realspace}
w(\vec{x}_0) = \sum_k \psi (\vec{x}_k-\vec{x}_0) I_p(\vec{x}_k)
\end{equation}
\noindent where $\vec{x}_0$ is the position of the considered source,
$I_p(\vec{x}_k)$ is the intensity of the pixel of the unfiltered image
located at the position $\vec{x}_k$ and $\psi$ is the kernel of the
filter. It can be shown that if $\psi=\psi_{MF}$ (that is, for the
matched filter) $w(\vec{x}_0)$ is the best possible linear estimator
(statistically unbiased and of maximum efficiency) of the flux of the
source. In particular, for the case of Gaussian noise $w(\vec{x}_0)$
is the maximum likelihood estimator of the flux of the source. As
shown in \citet{can06} the flux estimation when $\psi=\psi_2$ (that
is, with the MHW2) is comparable to that obtained with the MF.

The method adopted here differs from the one used by the WMAP
team. They have used the MF to detect point sources above the
$5\sigma$ level in the filtered images (full-sky maps, in their case),
but not to estimate their fluxes. These are derived by fitting, in the
unfiltered image, the pixel intensities around the point source to a
Gaussian profile plus a plane baseline. However, as already noted, the
profile of the source matches that of the beam, and is therefore
non--Gaussian. In this paper the true source profiles are used at
every step, up to the final flux unit conversions.

For each source we calculate the dispersion $\sigma$ as the square
root of the variance of the background pixels that are close to
the target source but are not ``contaminated'' by it. This number
can be easily inflated in the presence of contamination by nearby
sources or large scale structures that, in some cases, cannot be
removed completely by filtering. In order to avoid this, we first
select a shell of pixels around the source, with an inner radius
equal to the FWHM of the beam. This guarantees that the source
flux has decreased well below the background level. Subsequently,
we choose an outer radius encompassing a sufficiently large number
of pixels ($\sim 3000$) to give an accurate estimate of $\sigma$.
As a final step, we divide this shell in sectors -- 12, normally
-- and calculate their mean and dispersion. Strong contamination,
due to other sources present in the annulus, shows up as a
significant difference between the mean and the median; whenever
this difference exceeds two times the dispersion we excluded the
pixels in such sector from the calculation of the variance.

The mean values of $\sigma$, the rms error on the EPS fluxes, turn out
to be: 198, 231, 222, 255 and 399 mJy at 23, 33, 41, 61 and 94 GHz,
respectively. The variation of $\sigma$ with the WMAP channel is
determined by the beam shapes, the spectral behavior of the foreground
emission, the instrumental noise, etc.. It may be noted that these
uncertainties are typically 2 or 3 times higher than the uncertainties
quoted by Hinshaw et al.  (2006), which are probably underestimated,
as confirmed by the fact that their source counts show clear signs of
incompleteness at flux densities well above $5$ times their typical
errors.

\subsection{Bayesian correction to the fluxes} \label{sec:bayes}

Flux-limited surveys are affected by the well-known Eddington bias
\citep{edding} that leads to an overestimate of the flux of faint
sources. Hogg \& Turner (1998) have shown how to correct for this
effect if the underlying distribution of source fluxes is
known. Unfortunately, our knowledge of this distribution in the
frequency range covered by WMAP is rather poor. However, as shown in
the Appendix~\ref{app:bayes}, given a set of observed fluxes and the
associated values of the rms noise it is possible to simultaneously
estimate the slope of the flux distribution (i.e., of the differential
number counts) and to obtain an unbiased estimate of the source
fluxes. We have applied this method, based on the Bayesian approach
introduced by \citet{yo06}, to correct for the Eddington bias. The
estimated slopes of the differential number counts are 2.11, 2.34,
2.16, 2.14, and 2.16 for the 23, 33, 41, 61, and 94 GHz channels,
respectively, in very good agreement with the results of the ATCA 18
GHz survey (Ricci et al. 2004: $2.2\pm 0.2$), of the 9C survey at 15
GHz (Waldram et al. 2003: $2.15$), and of the 33 GHz VSA survey
(Cleary et al. 2005: $2.34^{+0.25}_{-0.26}$).

\section{The New Extragalactic WMAP Point Source (NEWPS) catalogue}  \label{sec:sims}

\subsection{Comparison between MHW2 and MF}

\begin{figure}
  \begin{center}
\includegraphics[width=0.45\textwidth]{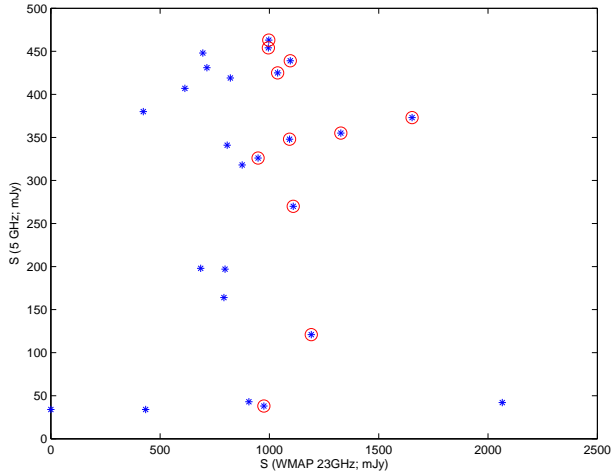}
\caption{5 GHz vs 23 GHz WMAP flux densities of the 25 sources missed
by our low-frequency selection. The source with zero 23 GHz flux was
not detected at 23 GHz, but at least at one higher frequency. Encircled dots
are sources that we recover at $\ge 5\sigma$.}\label{fig:missing_ps}
  \end{center}
\end{figure}

\begin{figure*}
\begin{center}
\includegraphics[width=0.95\textwidth]{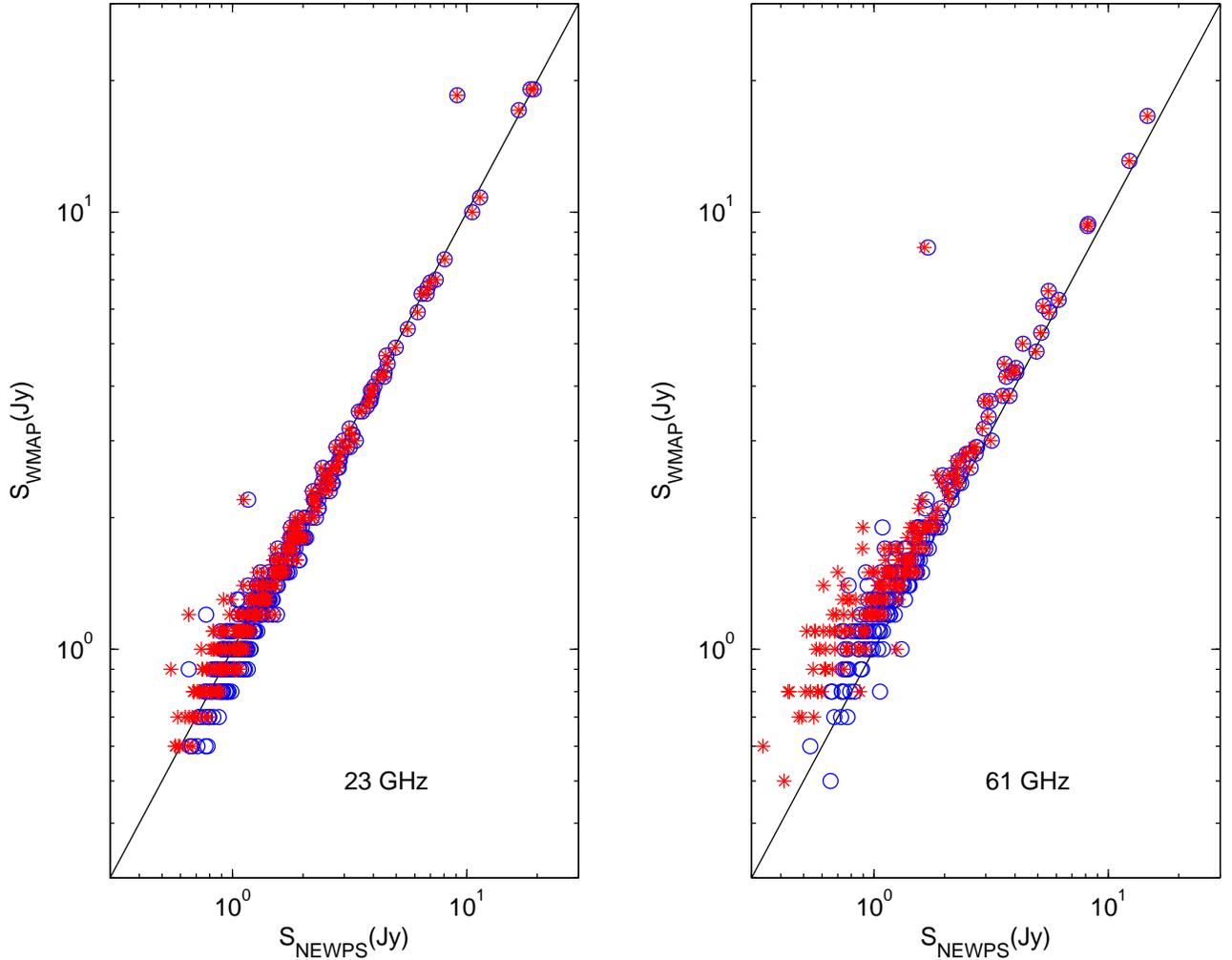}
\caption{Comparison of $NEWPS_{3\sigma}$ with WMAP flux densities at
23 and 61 GHz: \emph{Blue circles} show our uncorrected fluxes and
\emph{red asterisks} the corrected ones.}\label{fig:S_S}
\end{center}
\end{figure*}

\begin{figure}
  \begin{center}
    \includegraphics[width=0.45\textwidth]{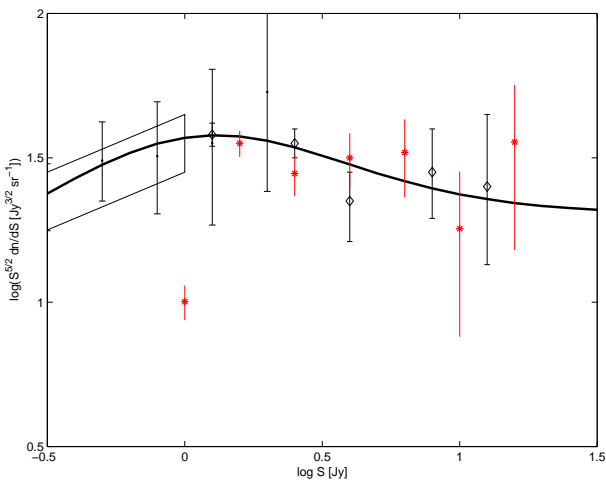}
    \caption{Number counts multiplied by $S^{5/2}$.
 \emph{Red asterisks}: counts from the $NEWPS_{5\sigma}$ catalogue at 33
      GHz (see text for more details). \emph{Black diamonds}:
      WMAP counts \citep{hinshaw06}.  \emph{Black dots}: ATCA 18 GHz
      pilot survey counts \citep{ricci04}. The parallelogram is from the DASI
      experiment at 31 GHz \citep{kov02}. The solid curve shows, for comparison, the counts
      predicted by the model by
      \citet{dz05}.}\label{fig:counts_33GHz}
  \end{center}
\end{figure}

We have carried out the detection/flux estimation process using the
two different filters, MHW2 and MF, previously discussed. We have
found that filtering with the MF introduces ringing effects around the
target EPS in at least 15\% of the images analyzed (see Figure
\ref{ringing}). These effects are stronger for the brightest
sources. Rings appear due to strong oscillations in the shape of the
matched filter which are determined by sharp features in the angular
power spectrum of temperature anisotropies in the sky patch. These
sharp features are likely to appear in regions showing a high
background signal and/or in regions where point sources constitute a
major component of the total intensity of the image. If the number of
images were small, we could check the images one--by--one visually and
study separately the anomalous cases, but this procedure is
impractical in the present context. With a {\it blind} approach we get
a number of spurious sources, mainly due to positive interferences
between rings. In the case of our non-blind approach rings may also
contaminate surrounding pixels, thus affecting the error estimates and
the flux estimates of the surrounding sources. On the contrary, we
found that the MHW2 filter does not introduce ringing effects in any
of the considered images.

The number of source detections above a certain $n\sigma$ threshold
obviously depends on the correct estimation of the background noise
level. Ringing effects affect negatively such estimate. We find that,
on average, by filtering with the MHW2 we correctly identify $\sim7\%$
more sources than with the MF\footnote{This percentage varies depending on the
observation frequency, from $\sim14\%$ to $\sim4\%$.}. For this reason
we decided to build the final catalogue using the detections/flux
estimates obtained with the MHW2.

\subsection{$NEWPS_{3\sigma}$}

To define our first SubCatalogue (SC), $NEWPS_{3\sigma}$, of sources
at $\geq 3\sigma$ we need to take into account that the IC was
generated from surveys with angular resolution $\le 4'.2$, i.e. much
higher than the WMAP resolution, so that we may have multiple IC
sources within one WMAP beam. In some cases the WMAP signal is
entirely due to the brightest source within the beam, whose flux,
attenuated by the response function, accounts also for the signal
detected in the direction of nearby sources. The latter are therefore
false detections that we have removed from the sample. A search within
circles of radius equal to 2 WMAP beam sizes (beam
size=FWHM$/2\sqrt{2\log(2)}$) centered on the position of each source,
starting from the brightest ones, has yielded numbers of false
detections decreasing from 354 at 23 GHz to 104 at 94 GHz.

The contamination of very bright sources can actually extend
beyond 2 WMAP beam sizes. We have therefore repeated the above
search up to 4 beam sizes, and removed those sources for which the
contamination by the bright source accounts for more than 50\% of
the detected signal. The number of such cases oscillate around 7
at all channels. In 13\% of the cases, the data are consistent
with more than one IC source contributing significantly to the
WMAP flux.

After having removed the 4 $|b|> 5^\circ$ sources known to be Galactic
(Taurus A, Orion A \& B, and IC\,418), we are left with
760(738+22 missed sources, WMAP sources that are not included in our
IC), 565(547+18), 536(518+18), 366(360+5) and 103(101+2) EPS with
signal-to-noise ratios $>3$ in the 23, 33, 41, 61, and 94 GHz
channels, respectively (see Table \ref{tb:summary0}). Finally, a cross
correlation of the catalogues at the different frequencies shows that
we have 933 (908+25) different EPS with signal-to-noise ratios $>3$ in
at least one WMAP channel, which constitutes our $NEWPS_{3\sigma}$
Catalogue.\footnote{All the catalogues can be obtained from
max.ifca.unican.es/caniego/NEWPS}

\subsection{$NEWPS_{5\sigma}$: A robust subsample}

Of the 381 sources that we detected at $\ge 5\sigma$
($NEWPS_{5\sigma}$ Catalogue in Table 4, see Appendix C., plus 12 EPS
detected at $>5\sigma$ listed in Table 5) only 283 are listed in the WMAP
3-yr catalogue (see Table \ref{tb:statistic} for a better
understanding of the detection statistics). As expected, the a priori
knowledge of source positions has allowed us to significantly increase
the detection efficiency. Additionally, in our approach, 39 (26+13)
WMAP sources present a signal-to-noise ratio $<5$ (as shown in Table
3) \footnote{However, 3 of the 26 WMAP sources with low frequency
counterparts included in the IC are not 5$\sigma$ detections in the
original WMAP catalogue.  These sources are: GB6 J1228+1124, GB6
J1231+1344 and GB6 J1439+4958.}.

Regarding the reliability of the detected sources near the $5\sigma$
threshold, we would expect that most of them, if not all, are actually
real $5\sigma$ sources. Since we have followed a non-blind approach we
know that there is a real source in every position we have considered,
but it is possible that a very weak source appears above this
threshold because it has been contaminated, e.g. by Galactic
foregrounds, CMB, etc.  In order to check this possibility we have
created a catalogue of false positions and we have repeated the
process of rotation, projection, filtering of each patch and flux
estimation exactly in the same way as we have done for detecting the
sources in $NEWPS_{5\sigma}$. Then we have counted the number of
$>5\sigma$ detections.  This number gives an idea of how likely is
that the foregrounds produce a spurious $>5\sigma$ detection and
therefore how reliable is our catalog for that threshold.

This catalogue of false positions is obtained by shifting 10 degrees
in the longitude coordinate the position in the sky of each source
outside $|b|>5^\circ$ in the IC. We have performed this test only at
23 GHz, because at this frequency we have the highest number of
detections and the lowest mean value of rms noise. As a result, we
have found $3\%$ of spurious $>5\sigma$ detections in the analysed
patches, which is in agreement with the numbers discussed by
\citet{ben03b} and \citet{hinshaw06}.

In addition, as already mentioned, there are 25 WMAP sources left out
by our low-frequency selection, but included in our analysis (see
Table 4). Only 12 of them are detected at $\ge 5\sigma$ by our
approach, and only 3 have 23 GHz flux densities above the estimated
completeness limit of 1.1 Jy (see Fig.~\ref{fig:counts_33GHz}).  Of
our 12 sources which are $\ge 5\sigma$ detections, 10 have
low-frequency flux densities $\ge 270\,$mJy (8 of them are above 340
mJy) and may well be variable sources, that happened to be in a
particularly `high' phase at the time of WMAP observations. The source
with $S_{5{\rm GHz}} = 120\,$mJy has an inverted spectrum (i.e. a
spectrum rising with frequency) and the last one, with $S_{5{\rm GHz}}
= 38\,$mJy, is a strong candidate to be a spurious source.

Finally, by exploiting the distribution of spectral indexes
$\alpha_5^{23}$, we could also tentatively estimate how many more
$5\sigma$ detections we missed at 23 GHz because of the adopted 5 GHz
flux limit $S_{5GHz}\geq 500$ mJy. From the observationally determined
number counts at 5 GHz (and the associated uncertainties) we estimate
that there should be $\sim$2100-2500 sources in the flux range $300<
S_{5GHz}< 500$ mJy, in the same sky area in which we find 2491 sources
with $S\geq 500$ mJy. For being detectable at 23 GHz, these sources
should show an inverted spectrum with slope $\alpha \le -0.63$. The
distribution of the spectral indexes $\alpha_5^{23}$ of sources in the
IC shows that only $\leq 1$\% of them show spectra so inverted.
This implies that we may expect $\leq 20$--25 sources with $300<
S_{5GHz}< 500$ to be detectable at $\ge 5\sigma$ at 23 GHz. The number
of expected detections decreases to $\le 5$ for $100< S_{5GHz}< 300$,
and is essentially zero at still fainter 5 GHz fluxes.  These
estimates are comparable with the number (11) of sources at 23 GHz
(see Table 3) detected by WMAP but missed by our selection criterion.
Note that, at higher frequencies, the numbers of missed detections,
which is more difficult to estimate due to the higher errors in flux,
decreases in parallel with the decrease of the number of detections
(see Table 2 and Section 5.1).

\section{Results and discussion} \label{sec:results}

\begin{deluxetable*}{cccrcrcl}
\tabletypesize{\scriptsize}
\tablewidth{0pt}
\tablecaption{Summary of the large-area surveys of point sources used in this work. \label{tb:summary}}%
 \tablehead{
   \textbf{Frequency} & \textbf{Catalogue} & \textbf{$S_{lim}$(mJy)}&
    \multicolumn{3}{c}{\textbf{DEC range}} & \textbf{Angular resolution} }
    \startdata &
    \textbf{GB6} & 18 & 0\phantom{5}\ & ---\ & +75\phantom{.5} & 3.5' & \citet{gre96} \\ &
    \textbf{PMNE} & 40 & -9.5\ & ---\ & +10\phantom{.5} & 4.2' & \citet{gri95}\\ 4.85 GHz &
    \textbf{PMNT} & 42 & -29\phantom{.5}\ & ---\ & -9.5 & 4.2' & \citet{gri94}\\ &
    \textbf{PMNZ} & 72 & -37\phantom{.5}\ & ---\ & -29\phantom{.5} & 4.2' & \citet{wri96}\\ &
    \textbf{PMNS} & 20 & -87.5\ & ---\ & -37\phantom{.5} & 4.2'& \citet{wri94}\\ \hline 1.4
    GHz & \textbf{NVSS} & 2.5 & -40\phantom{.5}\ & ---\ & +90\phantom{.5} & 45'' & \citet{con98}\\
    \hline 0.843 GHz & \textbf{SUMSS} & 18 & -50\phantom{.5}\ & ---\ & -30\phantom{.5} & 45''$\hbox{cosec}|\delta|$&\citet{mau03}\\
    & & 8 & -90\phantom{.5}\ & ---\ & -50\phantom{.5} & 45''$\hbox{cosec}|\delta|$&\\
    \enddata
\end{deluxetable*}

\begin{deluxetable*}{ccccccc}
\tabletypesize{\scriptsize}
\tablewidth{0pt}
\tablecaption{Summary of the results. \label{tb:summary0}}
 \tablehead{ & \textbf{23 GHz} & \textbf{33 GHz} & \textbf{41 GHz} & \textbf{61 GHz} & \textbf{94 GHz} &\textbf{Total} }
    \startdata
    \textbf{$\textless$ rms $\textgreater$ } & 198 & 231 & 222 & 255 & 399 & -- \\
    \textbf{$\alpha$}  & 2.11 & 2.34 & 2.16 & 2.14 & 2.16 & --\\
    \textbf{$NEWPS_{3\sigma}$} & 760 & 565 & 536 & 366 & 103 & 933 \\
                          & (738+22) & (547+18) & (518+18) & (360+6) & (101+2) & (907+25) \\
    \textbf{$NEWPS_{5\sigma}$} & 350 & 224 & 218 & 136 & 22 & 381\\
                          & (339+11) & (219+5) & (215+3) & (135+1) & (22+0) & (369+12)\\
    \textbf{WMAP 5$\sigma$} & 314 & 292 & 280 & 154 & 29 & 323\\ \enddata
    \tablecomments{The ``rms'' of the patches is in units of mJy. The
    slope $\alpha$ corresponds to the estimated slope of the flux
    distribution (i.e. of the differential number counts, dN/dS $\propto S^{-\alpha}$ ). For the
    $NEWPS_{5\sigma}$ and $NEWPS_{3\sigma}$ we show in parenthesis the
    number of detections coming from the initial catalogue ``IC'' and
    the number of detections among the 25 WMAP objects not present in IC (missed sources, see Table 4).}
\end{deluxetable*}

\begin{deluxetable*}{lcccc}
\tabletypesize{\scriptsize} \tablewidth{0pt} \tablecolumns{5}
\tablecaption{Detection statistics}
\tablehead{ & (All $\nu$) & (All $\nu$)& (23 GHz) & (23 GHz) \\
SubCatalogue (SC): & $3\sigma$ & $5\sigma$ & $5\sigma$ & $\geq$
$S_{lim}$ 1.1Jy} \startdata Number of WMAP EPS $\in$ SC & 297 & 271 &
258 & 186\\
Number of WMAP EPS $\notin$ SC & --- & 26 & 39 & 111\\
Number of Missed EPS $\in$ SC & 25 & 12 & 11 & 3\\
Number of Missed EPS $\notin$ SC & --- & 13 & 14 & 22\\
Dropped WMAP EPS & 1 & 1 & 1 &1\\
New EPS ($\notin$ in WMAP) & 611 & 98 & 81 & 43\\
\enddata
\tablecomments{Distribution of the detected sources in different
subsamples. The description has been made for 4 classes of EPS: WMAP
EPS (WMAP sources that appeared in our IC), Missed EPS (WMAP EPS
NOT included in our IC), Dropped EPS (the planetary nebula IC418
(PMNJ0527-1241)) and New EPS (detected EPS NOT included in the WMAP
EPS catalogue). The total number of sources in each subCatalogue,
detected by the MHW2 filter, are given by the sum of the entries in
lines (1)+(3)+(6), in agreement with the numbers in Table 2.}
\label{tb:statistic}
\end{deluxetable*}

\subsection{$NEWPS_{5\sigma}$ versus WMAP sources}

The WMAP 3-year catalogue (Hinshaw et al. 2006) lists 314, 292, 280,
154, and 29 $\ge 5\sigma$ detections at 23, 33, 41, 61, and 94 GHz,
respectively, to be compared with 350 (339+11 missed sources), 224
(219+5), 218 (215+3), 136 (135+1) and 22 sources in the
$NEWPS_{5\sigma}$ catalogue (see Table \ref{tb:summary0}). Although
the blind WMAP technique yielded a lower number of detections at 23
GHz and a lower total number, it was apparently more successful in
retrieving sources at $\ge 5\sigma$ in different channels. This may
be, to some extent, related to their lower error estimates. However,
note that the same sources can be detected in different
channels. Therefore, the total number is the intersection of the
channels, giving a global a reduced number for the WMAP case (323
vs. 381).

In Fig.~\ref{fig:S_S} we compare our flux estimates for sources in the
$NEWPS_{3\sigma}$ catalogue with the WMAP ones at 2 frequencies (23
and 61 GHz). We have plotted the corrected fluxes (\emph{red
asterisks}) as well as the uncorrected ones (\emph{blue circles}) to
show the effect of the Bayesian correction. At 23 GHz the agreement is
generally good and the correction makes no difference for fluxes
$\gtrsim 1.5\,$Jy, due to the fact that almost all the sources are
detected at high signal to noise ratios.

The most striking difference is found for \object{Fornax A}
(PMNJ0321-3658), represented by the isolated point on the top
right-hand corner of the 23 GHz channel and at the center of the upper
part of the 61 GHz panel. Our procedure yields 23 GHz and 61 GHz flux
densities of $9.1\pm 0.2$ and $1.6\pm 0.2$ Jy, to be compared with
$18.5\pm 3.6$ and $8.3\pm 2.1$ Jy, respectively, given in the WMAP
catalogue (fluxes obtained by aperture photometry). The problem here
is that Fornax A has a relatively weak core and 2 big lobes extending,
altogether, over $\simeq 1^\circ$, so that it cannot be treated as a
point source, even at the relatively low resolution of WMAP. As both
flux estimates are not reliable we have applied the NRAO Astronomical
Image Processing System (AIPS) software to the unfiltered patches,
fitting the source to a gaussian. The values obtained by this method
are $11.49\pm 0.63$ and $0.92\pm 0.72$ Jy, with a best fit
major(minor) axis of $1.18^\circ(0.81^\circ)$. Resolution effects may
be worrisome also for \object{Centaurus A} (PMNJ1325-4257), the
largest radio source in the sky at low frequencies, which is however
not present in the WMAP catalog. Therefore for this source too we have
checked our flux estimates with those obtained from AIPS, which gave
$48.12\pm 0.04$ and $20.31\pm 0.72$ Jy, for 23 GHz and 61 GHz
respectively, and a best fit major (minor) axis of
$1.13^\circ(0.84^\circ)$ for 23 GHz, in good agreement with our
results.

The second discrepant point (circle not far from the panel center at
23 GHz) is \object{PMN J0428-3756} for which we get a 23 GHz flux of
1.12 Jy, while the WMAP flux is 2.20 Jy. This source is an ATCA flux
calibrator. Its light
curve\footnote{www.narrabri.atnf.csiro.au/cgi-bin/Calibrators/calfhis.cgi?
source=0426-380\&band=12mm} at 20 GHz shows its flux increasing from
$\simeq 1\,$Jy in 2002 to a maximum of 2.04 Jy in 2004, and steadily
decreasing afterwards back to $\simeq 1\,$Jy in 2006.

The asterisks in the lower left-hand part of each panel correspond
to objects that we detect at $< 5\sigma$ at 23 GHz, whose flux
density is substantially decreased by the correction for the
Eddington bias.

At 61 GHz WMAP flux densities are systematically brighter than
ours by $\simeq 10\%$. This may be due to the use by the WMAP team
of Gaussian source profiles while, as discussed by
\citet{hinshaw06}, beams increasingly deviate from Gaussianity
with increasing frequency. On the other hand, we have taken into
account real symmetrized beam profiles. At this frequency many
sources fainter than $\simeq 2\,$Jy are detected by our procedure
at $< 5\sigma$ and have substantial corrections for the Eddington
bias, moving them to the left of the diagram.

\subsection{Source number counts}\label{sec:counts}

In Figure \ref{fig:counts_33GHz} we compare the number counts derived
from the $NEWPS_{5\sigma}$ catalogue at 33 GHz (red asterisks with Poisson
error bars) with other sets of observational data, specified in the
caption, and with the prediction of the model by \citet{dz05}. The
agreement is clearly good. The completeness flux limit of our
catalogue is around $\sim$1.1 Jy, as expected from the average value
of the flux error at this frequency, $\simeq 230\,$mJy.

Hinshaw et al. (2006) estimated the contribution of point sources
below the detection threshold to the anisotropy power spectrum at
40.7 GHz to be $A=0.017\pm 0.002\,\mu\hbox{K}^2$. Again, this is
in good agreement with the De Zotti et al. (2005) model, yielding,
at this frequency, $A=0.018\,\mu\hbox{K}^2$.

\section{Conclusions} \label{sec:Conclusions}

We have used the MHW2 filter (Gonz\'alez-Nuevo et al. 2006) to obtain
estimates of (or upper limits on) the flux densities at the WMAP
frequencies of a complete all-sky sample of 2491 sources at
$|b|>5^\circ$, brighter than 500 mJy at 5 GHz (or at 1.4 or 0.84 GHz,
in regions not covered by 5 GHz surveys but covered by either the NVSS
or the SUMSS). We have shown that the MHW2 filter has an efficiency
very similar to the MF and is much easier to use.

We have obtained flux density estimates for the 933 sources detected
at $\geq 3\sigma$, our $NEWPS_{3\sigma}$ Catalogue (see Section
4.2). Of the 381, presumably extragalactic, sources detected at $\geq
5\sigma$, 369 of which constitute our $NEWPS_{5\sigma}$ Catalogue
(Section 4.3), 98 (i.e. 26\%) are ``new'', in the sense that they are
not present in the WMAP catalogue; 43 of them are above the estimated
completeness limit of the WMAP survey: $\simeq 1.1$ Jy at 23 GHz. This
illustrates how the prior knowledge of source positions can help their
flux measurements.

On the other hand, 39 (23+13) WMAP extragalactic sources with
low-frequency counterparts included in our sample were detected by us
at $< 5\sigma$. This is probably due to the fact that our error
estimates exceed substantially those given by the WMAP team,
particularly at 23 GHz, where we have the highest detection rate. In
fact, the WMAP errors do not correspond to the rms fluctuations in the
source neighborhood but to the uncertainties in the amplitude of the
Gaussian fit. As a consequence, the WMAP catalogue starts being
incomplete at flux densities more than 2 times higher than 5 times
their typical formal errors.

Our flux density estimates for sources detected at $\ge 5\sigma$
are generally in very good agreement with the WMAP ones at 23 GHz.
At higher frequencies WMAP fluxes tend to be slightly but
systematically higher than ours, probably due to having ignored
the deviations, increasing with frequency, of the point spread
function from a Gaussian shape. Our estimates use the real beam
shape at every frequency. For only one source we have a strong
discrepancy with WMAP. Such source, Fornax A, is known to have
powerful lobes extending over $\sim 1^\circ$, and is therefore
resolved by WMAP. Thus, the point source assumption on which both
WMAP and our flux estimates rely, is clearly not appropriate and
yields unreliable results. A smaller, but still significant,
discrepancy, is found for PMN J0428-3756.

We have also worked out and applied a method to correct flux
estimates for the Eddington bias, without the need of an {\it
a-priori} knowledge of the slope of source counts below the
detection limit.

Our selection criterion leaves out 25 WMAP sources, only 12 of which
however turn out to be $\ge 5\sigma$ detections after our analysis,
and only 3 have 23 GHz fluxes $\gtrsim 1.1\,$Jy, the estimated
completeness limit of the survey. Thus, our approach has proven to be
competitive with, and complementary to the blind technique adopted by
the WMAP team. In fact we missed 3 sources brighter than the estimated
completeness limit ($S_{23\rm GHz} = 1.1\,$Jy), but we detected 42 new
ones. 

On the whole, 26\% of sources we have detected at $\ge 5\sigma$ are
not present in the WMAP catalogue. On the other hand, the efficiency
of the process is low. Only 381 of the 2491 sources in our input
sample were detected at $\ge 5\sigma$ in at least one WMAP channel.

In a forthcoming paper, we will exploit our catalogue to investigate
the high-frequency properties of sources selected at low frequencies.

\section*{acknowledgements}
We acknowledge partial financial support from the Spanish Ministry of
Education (MEC) under project ESP2004--07067--C03--01. MLC acknowledge
a FPI fellowship of the Spanish Ministry of Education and Science
(MEC). JGN acknowledges a postdoctoral position at the SISSA-ISAS
(Trieste). We are grateful to Corrado Trigilio for help with the AIPS
package and N. Odegard for clarifying some aspects of the WMAP EPS
catalogue.

\appendix

\section[]{Linear filters in two dimensions} \label{app:filters}

The approach adopted in this paper assumes that in the center of
each image (sky patch) there is a point source with unknown flux
density, $A$. As usual we describe the source as
\begin{equation} \label{eq:source}
s(\vec{x})=A\tau(\vec{x}),
\end{equation}
\noindent where $\tau(\vec{x})$ is the source \emph{profile}. We
will assume circular symmetry, so that $\tau(\vec{x})=\tau(x)$,
$x=|\vec{x}|$. For point sources the profile is equal to the beam
response function of the detector. For a circular Gaussian beam we
have
\begin{equation} \label{eq:gauss_source}
\tau (x) = e^{- x^2/2R_s^2},
\end{equation}
\noindent where $R_s$ is the angular radius of the beam response
function. In the Fourier space the beam profile of
eq.~(\ref{eq:gauss_source}) is
\begin{equation} \label{eq:gauss_source_fourier}
\tau (q) = e^{- q^2 R_s^2/2},
\end{equation}
\noindent where $q\equiv|\vec{q}|$. Actually the WMAP beams are
not Gaussian and we will use the real symmetrized radial beam
profiles for the different WMAP channels to construct our filters.

Let us consider a 2D--filter $\Psi (\vec{x}; R, \vec{b})$, where
$R$ and $\vec{b}$ define a scaling and a translation respectively.
Then
\begin{equation} \label{eq:psi}
  \Psi(\vec{x}; R, \vec{b}) \equiv \frac{1}{R^2} \psi \left(
  \frac{|\vec{x} - \vec{b}|}{R} \right) .
\end{equation}
If we filter our field $f(\vec{x})$ with  $\Psi (\vec{x}; R,
\vec{b})$, the filtered map will be
\begin{equation} \label{eq:wav}
  w(R, \vec{b}) = \int d\vec{x}\,f(\vec{x})\Psi (\vec{x}; R, \vec{b}).
\end{equation}
\noindent The filter normalization preserves the amplitude of the
source at its position ($b=0$) after filtering:
\begin{equation}
  \int d\vec{x} \, \tau(\vec{x}) \Psi(\vec{x}; R, \vec{0}) = 1.
\end{equation}
\noindent The n-th moment of the filtered map is defined as
\begin{equation}
  \sigma_n^2 \equiv 2 \pi \int_0^{\infty} dq \ q^{1+2n} P(q)
  \psi^2(q).
\end{equation}
\noindent where $P(q)$ is the power spectrum of the unfiltered map
and $\psi(q)$ is the Fourier transform of $\psi(x)$:
\begin{equation}
  \label{cc2}
  \psi(q) = \int_0^{\infty}dx\,xJ_0(qx)\psi (x).
\end{equation}
\noindent Here $\vec{q}$ is the wave number, $q\equiv |\vec{q}|$
and $J_0$ is the Bessel function of the first kind. The
zeroth-order moment, $\sigma_0$, which is the dispersion of the
filtered map, will be denoted, for simplicity, as  $\sigma$. It is
a function of the filter scaling $R$, of the source profile $\tau$
and of the power spectrum $P(q)$ of the unfiltered image. Then,
for a given image and a given source profile it is possible to
optimize the ratio $A/\sigma$ just by modifying the scaling $R$.
The scale $R_0$ at which $A/\sigma$ is maximum is called
\emph{optimal scale}. It has been shown
\citep[]{patri01a,patri03,can04,can05} that, working at this scale
at which a filter maximizes the flux of a compact source with
respect to the average surrounding background, the filter
performance in detecting point sources and estimating their flux
is maximized.

\section[]{Derivation of the Bayesian correction formulae}
  \label{app:bayes}


\subsection{The Eddington bias}

The distribution, normalized to unity, of true fluxes, $S$, of
extragalactic sources, is usually well described by a power law
\begin{equation} \label{eq:powlaw}
P(S|q) = k S^{-(1+q)}, \ \ \ S \geq S_m
\end{equation}
\noindent where the normalization $k$ is
\begin{equation} \label{eq:normalization}
k=q {S_m}^q, \ \ \ q > 0.
\end{equation}
The observed fluxes are contaminated by noise. Let $\mathbf{S^o} =
\lbrace S^o_1, \ldots, S^o_N \rbrace$ be the observed fluxed of
the $N$ galaxies detected above a given flux threshold and
$\mathbf{S} = \lbrace S_1, \ldots, S_N \rbrace$ their true fluxes.
In the case of Gaussian noise we have
\begin{equation} \label{eq:noise1}
P \left(\mathbf{S^o} \big| \mathbf{S},q\right) \propto
\mathrm{exp} \left[ -\frac{1}{2} \sum_{i=1}^{N}
\left(\frac{S^o_i-S_i}{\sigma_i^2} \right)^2 \right]\ ,
\end{equation}
\noindent where $\sigma_i$ is the rms noise for the $i$th source.
Since we select only sources above a certain threshold, a
selection effect appears: sources whose intrinsic flux is lower
than the detection threshold may be detected due to positive noise
fluctuations and, on the other hand, sources that should be
detected because their flux is greater than the detection
threshold may be missed due to negative fluctuations. Since
according to eq.~(\ref{eq:powlaw}) there are more faint than
bright sources we end up with an excess of sources whose flux is
overestimated. This is the \emph{Eddington bias} \citep{edding}.

\subsection{Bayesian slope determination and flux correction}

>From the Bayes' theorem we have
\begin{eqnarray} \label{eq:bayes}
  P\left(\mathbf{S},q \big| \mathbf{S^o}\right) & = &
  \frac{P\left(\mathbf{S^o} \big| \mathbf{S},q\right)
    P(\mathbf{S},q)}{P(\mathbf{S^o})} \nonumber \\ & \propto &
  P\left(\mathbf{S^o} \big| \mathbf{S},q\right) P(\mathbf{S}|q) P(q).
\end{eqnarray}
\noindent We will assume that $P(q)$ is uniform, that the source
fluxes take values drawn at random from the distribution of
eq.~(\ref{eq:powlaw}), and that the flux of any given source is
independent of the fluxes of the other sources. Therefore
\begin{equation}
P\left(\mathbf{S} \big| q \right) = \prod_{i=1}^N P(S_i|q) = q^N
S_m^{Nq} \left(\prod_{i=1}^N S_i \right)^{-(q+1)}.
\end{equation}
\noindent Therefore, using eqs. (\ref{eq:noise1}) and
(\ref{eq:bayes}) we have
\begin{eqnarray} \label{eq:P}
  P\left(\mathbf{S},q \big| \mathbf{S^o}\right) & \propto & \mathrm{exp}
  \left[ -\frac{1}{2} \sum_{i=1}^{N} \left(\frac{S^o_i-S_i}{\sigma_i^2}
    \right)^2 \right] \times \nonumber \\ & & q^N S_m^{Nq}
  \left(\prod_{i=1}^N S_i \right)^{-(q+1)}.
\end{eqnarray}
If the slope $q$ is known, the maximum likelihood estimator of the
fluxes of the sources is easily calculated (Hogg \& Turner 1998):
\begin{equation} \label{eq:MLES}
S_i = \frac{S^o_i}{2} \left[1+\sqrt{1-\frac{4(1+q)}{r_i^2}} \right], \
\ \ r_i \geq 2\sqrt{1+q},
\end{equation}
\noindent where $r_i=S^o_i/\sigma_i$ is the signal to noise ratio
of the source. Conversely, if the true intrinsic fluxes of the
sources $\mathbf{S}$ were known and the slope $q$ unknown, the
maximum likelihood estimator for the value of $q$ would be
\begin{equation} \label{eq:MLEq}
q=\left[ \frac{1}{N} \sum_{i=1}^{N} \mathrm{ln} \left( \frac{S_i}{S_m}
\right) \right]^{-1}.
\end{equation}
Unfortunately, in many cases, and particular in the case of WMAP
surveys, neither $q$ nor $\mathbf{S}$ are known a priori. Then it
is necessary to solve simultaneously for the two unknowns. A way
to do this is to introduce eq.~(\ref{eq:MLES}) into
eq.~(\ref{eq:MLEq}), which gives the implicit equation
\begin{equation} \label{eq:MLEimp}
\frac{1}{q} = \frac{1}{N} \sum_{i=1}^N \left[ \mathrm{ln} \left(
\frac{S^o_i}{S^o_m} \right) + \mathrm{ln} \left(
\frac{1+\Delta_i}{1+\Delta_m} \right) \right],
\end{equation}
\noindent where
\begin{equation}
\Delta_i = \sqrt{1-\frac{4(1+q)}{r_i^2}}.
\end{equation}
Equation (\ref{eq:MLEimp}) can be solved numerically if the
minimum signal to noise ratio of the galaxies considered satisfies
the condition $r_m^2 \geq 4(1+q)$. Once $q$ is estimated, the
fluxes $\mathbf{S}$ can be estimated using eq. (\ref{eq:MLES}).

The asymptotic limits of the estimators, valid in the high signal
to noise regime, that is, for $r_m \gg q$, are:
\begin{eqnarray} \label{eq:asym}
q & \simeq & \left[ \frac{1}{N}\sum_{i=1}^{N} \mathrm{ln} \left(
\frac{S^o_i}{S^o_m} \right) \right]^{-1} \\
S_i & \simeq & S^o_i \left[ 1- \frac{1+q}{r_i^2} \right].
\end{eqnarray}

\section[]{$NEWPS_{5\sigma}$ source catalogue}\label{app:neps5}

The $NEWPS_{5\sigma}$ Catalogue consists of 369 entries corresponding
to all the EPS detected in the WMAP 3-yr full-sky maps at the
$>5\sigma$ level, after filtering with the MHW2 as discussed in the
text. The 25 missed sources (WMAP detected sources that did not appear
in our IC) are listed in Table 4 while the 369 IC sources detected at
$>5\sigma$ are presented in Table 5. For each EPS in the Catalogue,
and from left to right, we list the following data: the Equatorial,
$\alpha$, $\delta$, coordinates of the center of the pixel; the NON
CORRECTED source fluxes\footnote{The catalogue with the corrected
fluxes can be obtained from max.ifca.unican.es/caniego/NEWPS}, $S_\nu$, in
each WMAP channel identified by the channel symbols (K, Ka, Q, V and
W) and their corresponding estimated errors ($\sigma$), in Jy; the
position in the 3-year WMAP catalogue; the closest source present in
the PMN or GB6 catalogues or the brightest NVSS or SUMSS source within
the resolution element (i.e., inside the FWHM of the beam of the given
WMAP frequency channel) and a ``M'' label that means that at least
another source inside the beam has a 5 GHz flux within a factor of 2
from the brightest one, that we list as the likely
identification. Table 4 has an additional column listing the 5 GHz
fluxes. For those sources not detected at $\geq 5\sigma$, we have used
``--'' in the corresponding channel column.

\begin{deluxetable*}{rrrrrrrrrlc} 
\tabletypesize{\scriptsize}
\tablewidth{0pt}
\tablecaption{Missed EPS}%
\tablehead{
  RA & Dec & K($\sigma$) & Ka($\sigma$) & Q($\sigma$) & V($\sigma$) & W($\sigma$)& S$_{5GHz}$
  &WMAP& low freq. id.& \\
  h & deg & Jy(Jy) & Jy(Jy) & Jy(Jy) & Jy(Jy) & Jy(Jy) & Jy & & & }
 \startdata
 0.434& -35.197&  1.19 (0.15)&   0.96 (0.21)&  1.25 (0.20)& 1.06 (0.26)& 0.82 (0.38)& 0.12&   6& PMNJ0026-3512  &   \\
 0.493&   5.922&  1.09 (0.16)&   1.43 (0.22)&  0.89 (0.21)& 0.74 (0.26)& 1.22 (0.45)& 0.35&   7& PMNJ0029+0554  &   \\
 5.229& -20.268&  0.71 (0.17)&   0.51 (0.20)&  0.50 (0.21)& 0.41 (0.23)& 0.25 (0.38)& 0.43&  68& PMNJ0514-2029  &  \\
 5.321&  -5.675&  2.07 (0.63)&   1.43 (0.82)&  1.06 (0.67)& 0.41 (0.45)& 0.15 (0.56)& 0.04&  71& PMNJ0520-0537  &  \\
 5.427& -48.449&  1.04 (0.16)&   1.25 (0.24)&  1.25 (0.28)& 0.63 (0.25)& 1.03 (0.40)& 0.42&  74& PMNJ0526-4830  &  \\
 5.680& -54.277&  1.65 (0.17)&   1.52 (0.20)&  1.79 (0.18)& 1.22 (0.21)& 0.50 (0.31)& 0.37&  77& PMNJ0540-5418  &   \\
 5.839& -57.551&  1.33 (0.18)&   0.94 (0.21)&  1.14 (0.19)& 0.54 (0.21)& 0.59 (0.34)& 0.35&  79& PMNJ0550-5732  &   \\
 6.003& -45.466&  0.61 (0.20)&   0.95 (0.21)&  0.60 (0.20)& 0.66 (0.23)& 0.21 (0.39)& 0.41&  81& PMNJ0559-4529  &   \\
 8.273& -24.423&  1.11 (0.18)&   0.73 (0.21)&  0.87 (0.21)& 0.64 (0.25)& 0.32 (0.40)& 0.27& 108& PMNJ0816-2421  &   \\
10.544&  41.306&  1.10 (0.18)&   1.02 (0.20)&  0.92 (0.21)& 0.87 (0.26)& 0.94 (0.42)& 0.44& 133& GB6 J1033+4115 &   \\
11.034& -44.011&  0.70 (0.19)&   0.92 (0.21)&  0.77 (0.20)& 0.35 (0.22)& 0.23 (0.34)& 0.45& 142& PMNJ1102-4404  &   \\
11.829& -79.554&  0.97 (0.18)&   0.69 (0.28)&  0.51 (0.24)& 0.41 (0.23)& 0.10 (0.41)& 0.04& 152& PMNJ1150-7918  &   \\
12.056&  48.104&  0.79 (0.24)&   0.67 (0.22)&  0.67 (0.23)& 0.62 (0.22)& 0.14 (0.34)& 0.16& 156& GB6 J1203+4803 &   \\
12.464&  11.406&  0.00 (0.00)&   0.86 (0.32)&  0.88 (0.24)& 0.53 (0.27)& 0.43 (0.56)& 0.03& 162& GB6 J1228+1124 &   \\
12.519&  13.857&  0.91 (0.29)&   0.49 (0.35)&  0.39 (0.25)& 0.76 (0.30)& 0.20 (0.54)& 0.04& 165& GB6 J1231+1344 &   \\
13.047&  48.947&  0.43 (0.18)&   0.64 (0.21)&  0.68 (0.18)& 0.61 (0.21)& 1.09 (0.31)& 0.03& 170& GB6 J1303+4848 &   \\
13.559&  27.396&  0.88 (0.21)&   0.81 (0.26)&  0.68 (0.27)& 0.52 (0.25)& 1.46 (0.35)& 0.32& 181& GB6 J1333+2725 &   \\
14.668&  49.973&  0.69 (0.18)&   1.01 (0.22)&  0.84 (0.20)& 0.45 (0.25)& 0.74 (0.33)& 0.20& 196& GB6 J1439+4958 &   \\
16.807&  41.239&  0.80 (0.22)&   0.71 (0.31)&  0.71 (0.27)& 0.29 (0.23)& 0.00 (0.38)& 0.20& 222& GB6 J1648+4104 &   \\
16.994&  68.495&  0.42 (0.17)&   0.64 (0.18)&  0.72 (0.16)& 0.79 (0.19)& 0.52 (0.29)& 0.38& 228& GB6 J1700+6830 &   \\
17.128&   1.786&  1.00 (0.20)&   0.93 (0.24)&  0.80 (0.19)& 0.76 (0.28)& 0.70 (0.45)& 0.46& 230& PMNJ1707+0148  &   \\
17.618& -79.578&  0.82 (0.18)&   0.78 (0.20)&  0.81 (0.18)& 0.74 (0.21)& 0.69 (0.38)& 0.42& 235& PMNJ1733-7935  &   \\
22.496& -20.853&  0.95 (0.13)&   0.83 (0.20)&  0.78 (0.23)& 0.70 (0.27)& 1.28 (0.47)& 0.33& 297& PMNJ2229-2049  &   \\
22.939& -20.191&  1.00 (0.20)&   0.61 (0.22)&  0.76 (0.24)& 0.80 (0.30)& 0.41 (0.45)& 0.45& 305& PMNJ2256-2011  &   \\
23.266& -50.286&  0.81 (0.18)&   1.12 (0.18)&  0.77 (0.19)& 0.80 (0.23)& 0.61 (0.35)& 0.34& 307& PMNJ2315-5018  &   \\
\enddata
\end{deluxetable*}

\LongTables 
\begin{deluxetable*}{rrrrrrrrlc}
\tabletypesize{\scriptsize}
\tablewidth{0pt}
\tablecaption{$NEWPS_{5\sigma}$ source catalogue}%
\tablehead{
  RA & Dec & K($\sigma$) & Ka($\sigma$) & Q($\sigma$) & V($\sigma$) & W($\sigma$)
  &WMAP& low freq. id.& \\
  h & deg & Jy(Jy) & Jy(Jy) & Jy(Jy) & Jy(Jy) & Jy(Jy) & & & }
 \startdata
 0.103&  -6.481&   2.7 (0.2)&   2.3 (0.2)&   2.4 (0.2)&   2.2 (0.3)&   --~~~~~&   1& PMNJ0006-0623        &   \\
 0.214& -39.957&   1.3 (0.2)&   1.2 (0.2)&   --~~~~~&   --~~~~~&   --~~~~~&   2& PMNJ0013-3954        &   \\
 0.326&  26.031&   1.1 (0.2)&   --~~~~~&   --~~~~~&   --~~~~~&   --~~~~~&   3& NVSS J001939+260245  & M \\
 0.328&  20.445&   1.0 (0.2)&   --~~~~~&   --~~~~~&   --~~~~~&   --~~~~~&   4& GB6 J0019+2021       &   \\
 0.424& -26.067&   1.1 (0.2)&   --~~~~~&   --~~~~~&   --~~~~~&   --~~~~~&   5& PMNJ0025-2602        & M \\
 0.789& -25.251&   1.2 (0.2)&   --~~~~~&   --~~~~~&   --~~~~~&   --~~~~~&   9& PMNJ0047-2517        &   \\
 0.790& -73.125&   1.9 (0.2)&   1.4 (0.2)&   1.5 (0.2)&   --~~~~~&   --~~~~~&    & PMNJ0047-7308        &   \\
 0.827& -57.637&   1.4 (0.2)&   1.2 (0.2)&   0.9 (0.2)&   --~~~~~&   --~~~~~&  10& PMNJ0050-5738        &   \\
 0.847&  -9.429&   1.1 (0.2)&   --~~~~~&   --~~~~~&   --~~~~~&   --~~~~~&  13& PMNJ0050-0928        &   \\
 0.854&  -6.828&   1.2 (0.2)&   --~~~~~&   --~~~~~&   --~~~~~&   --~~~~~&  11& PMNJ0051-0650        &   \\
 0.962&  30.412&   --~~~~~&   1.2 (0.2)&   --~~~~~&   --~~~~~&   --~~~~~&    & GB6 J0057+3021       &   \\
 0.987& -72.215&   2.3 (0.2)&   1.4 (0.2)&   1.4 (0.2)&   --~~~~~&   --~~~~~&    & PMNJ0059-7210        &   \\
 1.099&  48.322&   --~~~~~&   1.2 (0.2)&   --~~~~~&   --~~~~~&   --~~~~~&    & GB6 J0105+4819       &   \\
 1.115& -40.584&   1.8 (0.1)&   1.9 (0.2)&   1.7 (0.2)&   1.1 (0.2)&   --~~~~~&  14& PMNJ0106-4034        &   \\
 1.140&   1.583&   2.6 (0.2)&   2.3 (0.3)&   2.2 (0.2)&   2.0 (0.3)&   --~~~~~&  16& PMNJ0108+0134        &   \\
 1.152&  13.335&   1.4 (0.2)&   1.1 (0.2)&   --~~~~~&   --~~~~~&   --~~~~~&  15& GB6 J0108+1319       &   \\
 1.270& -11.686&   1.4 (0.2)&   1.1 (0.2)&   1.3 (0.2)&   1.4 (0.3)&   --~~~~~&  18& PMNJ0116-1136        &   \\
 1.318& -21.685&   0.8 (0.2)&   --~~~~~&   --~~~~~&   --~~~~~&   --~~~~~&    & PMNJ0118-2141        &   \\
 1.361&  11.807&   1.2 (0.2)&   --~~~~~&   1.1 (0.2)&   --~~~~~&   --~~~~~&  19& GB6 J0121+1149       &   \\
 1.379&  25.070&   1.0 (0.2)&   --~~~~~&   --~~~~~&   --~~~~~&   --~~~~~&    & GB6 J0122+2502       &   \\
 1.421&  -0.082&   1.2 (0.2)&   --~~~~~&   --~~~~~&   --~~~~~&   --~~~~~&  20& PMNJ0125-0005        &   \\
 1.544& -16.863&   1.8 (0.2)&   1.6 (0.2)&   1.8 (0.2)&   1.3 (0.2)&   --~~~~~&  21& PMNJ0132-1654        &   \\
 1.551& -52.009&   0.7 (0.1)&   --~~~~~&   --~~~~~&   --~~~~~&   --~~~~~&    & PMNJ0133-5159        &   \\
 1.624&  47.842&   4.4 (0.2)&   4.3 (0.2)&   4.0 (0.2)&   3.0 (0.3)&   --~~~~~&  22& GB6 J0136+4751       &   \\
 1.626& -24.473&   1.1 (0.2)&   1.1 (0.2)&   1.6 (0.2)&   1.4 (0.3)&   --~~~~~&  23& PMNJ0137-2430        &   \\
 1.824&   5.973&   0.9 (0.2)&   --~~~~~&   --~~~~~&   --~~~~~&   --~~~~~&    & PMNJ0149+0556        &   \\
 1.870&  22.043&   1.3 (0.2)&   1.6 (0.2)&   --~~~~~&   --~~~~~&   --~~~~~&  24& GB6 J0152+2206       &   \\
 2.082& -17.060&   0.8 (0.2)&   --~~~~~&   --~~~~~&   --~~~~~&   --~~~~~&    & PMNJ0204-1701        & M \\
 2.083&  15.271&   1.6 (0.2)&   1.5 (0.2)&   1.3 (0.2)&   --~~~~~&   --~~~~~&  25& GB6 J0204+1514       &   \\
 2.083&  32.225&   1.6 (0.2)&   1.4 (0.2)&   1.3 (0.2)&   --~~~~~&   --~~~~~&  26& GB6 J0205+3212       &   \\
 2.183& -51.009&   2.8 (0.2)&   2.9 (0.2)&   2.9 (0.2)&   2.6 (0.2)&   --~~~~~&  27& PMNJ0210-5101        &   \\
 2.299&  73.874&   1.9 (0.2)&   1.4 (0.2)&   1.3 (0.2)&   --~~~~~&   --~~~~~&    & GB6 J0217+7349       &   \\
 2.317&   1.394&   1.0 (0.2)&   --~~~~~&   --~~~~~&   --~~~~~&   --~~~~~&    & PMNJ0219+0120        &   \\
 2.355&  35.927&   1.2 (0.2)&   1.1 (0.2)&   --~~~~~&   --~~~~~&   --~~~~~&  28& GB6 J0221+3556       &   \\
 2.380& -34.726&   0.8 (0.1)&   --~~~~~&   --~~~~~&   --~~~~~&   --~~~~~&  29& PMNJ0222-3441        &   \\
 2.390&  43.035&   1.9 (0.2)&   1.3 (0.2)&   1.5 (0.2)&   --~~~~~&   --~~~~~&  30& GB6 J0223+4259       &   \\
 2.529&  13.302&   1.4 (0.2)&   --~~~~~&   --~~~~~&   --~~~~~&   --~~~~~&  31& GB6 J0231+1323       & M \\
 2.627&  28.805&   3.5 (0.2)&   2.9 (0.2)&   3.3 (0.3)&   2.7 (0.3)&   --~~~~~&  32& GB6 J0237+2848       &   \\
 2.644&  16.561&   1.5 (0.2)&   1.5 (0.3)&   1.5 (0.2)&   1.6 (0.3)&   --~~~~~&  33& GB6 J0238+1637       &   \\
 2.666&   4.250&   0.9 (0.2)&   --~~~~~&   --~~~~~&   --~~~~~&   --~~~~~&    & PMNJ0239+0416        &   \\
 2.858&  43.264&   1.1 (0.2)&   --~~~~~&   --~~~~~&   --~~~~~&   --~~~~~&    & GB6 J0251+4315       & M \\
 2.890& -54.680&   2.5 (0.1)&   2.6 (0.2)&   2.7 (0.2)&   2.1 (0.2)&   --~~~~~&  34& PMNJ0253-5441        &   \\
 2.993&  -0.342&   1.1 (0.2)&   --~~~~~&   --~~~~~&   --~~~~~&   --~~~~~&    & PMNJ0259-0020        &   \\
 3.065&  47.292&   1.3 (0.2)&   1.4 (0.2)&   --~~~~~&   --~~~~~&   --~~~~~&    & GB6 J0303+4716       &   \\
 3.068& -62.246&   1.4 (0.2)&   1.4 (0.2)&   1.3 (0.2)&   1.2 (0.2)&   --~~~~~&  35& PMNJ0303-6211        &   \\
 3.143&   4.099&   1.3 (0.2)&   --~~~~~&   --~~~~~&   --~~~~~&   --~~~~~&  36& PMNJ0308+0406        &   \\
 3.170& -60.972&   1.1 (0.2)&   1.3 (0.2)&   --~~~~~&   --~~~~~&   --~~~~~&  37& PMNJ0309-6058        & M \\
 3.196& -76.889&   1.2 (0.2)&   1.0 (0.2)&   --~~~~~&   --~~~~~&   --~~~~~&  38& PMNJ0311-7651        &   \\
 3.301&  41.831&   --~~~~~&   --~~~~~&   3.1 (0.2)&   2.0 (0.3)&   --~~~~~&    & GB6 J0318+4153       &   \\
 3.329&  41.530&  11.4 (0.2)&   8.3 (0.2)&   7.1 (0.2)&   4.9 (0.3)&   --~~~~~&  39& GB6 J0319+4130       &   \\
 3.361& -37.023&   9.1 (0.2)&   5.3 (0.2)&   3.5 (0.2)&   1.7 (0.2)&   --~~~~~&  40& PMNJ0321-3711        & M \\
 3.365&  12.347&   1.4 (0.2)&   --~~~~~&   --~~~~~&   --~~~~~&   --~~~~~&    & GB6 J0321+1221       &   \\
 3.407& -37.222&   --~~~~~&   4.8 (0.2)&   2.1 (0.2)&   --~~~~~&   --~~~~~&    & PMNJ0324-3716        &   \\
 3.425&  22.397&   1.1 (0.2)&   --~~~~~&   --~~~~~&   --~~~~~&   --~~~~~&  41& GB6 J0325+2223       & M \\
 3.499& -23.909&   1.0 (0.2)&   1.2 (0.2)&   1.1 (0.2)&   --~~~~~&   --~~~~~&  42& PMNJ0329-2357        &   \\
 3.573& -40.148&   1.6 (0.2)&   1.4 (0.2)&   1.4 (0.2)&   1.5 (0.2)&   --~~~~~&  43& PMNJ0334-4008        &   \\
 3.604&  32.255&   --~~~~~&   1.9 (0.3)&   2.1 (0.3)&   1.7 (0.3)&   --~~~~~&    & GB6 J0336+3218       &   \\
 3.606& -13.048&   1.0 (0.2)&   --~~~~~&   1.1 (0.2)&   --~~~~~&   --~~~~~&  44& PMNJ0336-1302        &   \\
 3.654&  -1.767&   2.8 (0.2)&   2.5 (0.3)&   2.7 (0.2)&   1.5 (0.3)&   2.6 (0.5)&  45& PMNJ0339-0146        &   \\
 3.678& -21.311&   1.2 (0.1)&   1.1 (0.2)&   1.1 (0.2)&   1.3 (0.2)&   --~~~~~&  46& PMNJ0340-2119        & M \\
 3.813& -27.772&   1.0 (0.2)&   --~~~~~&   --~~~~~&   1.1 (0.2)&   --~~~~~&  47& PMNJ0348-2749        & M \\
 3.985&  10.395&   1.4 (0.2)&   --~~~~~&   --~~~~~&   --~~~~~&   --~~~~~&  48& GB6 J0358+1026       &   \\
 4.069& -36.080&   4.0 (0.2)&   4.3 (0.2)&   4.2 (0.2)&   3.9 (0.2)&   2.6 (0.3)&  49& PMNJ0403-3605        &   \\
 4.096& -13.143&   1.9 (0.2)&   2.0 (0.2)&   1.6 (0.2)&   1.4 (0.3)&   --~~~~~&  50& PMNJ0405-1308        &   \\
 4.111& -38.449&   --~~~~~&   1.3 (0.3)&   --~~~~~&   --~~~~~&   --~~~~~&  51& PMNJ0406-3826        &   \\
 4.142& -75.112&   0.8 (0.2)&   --~~~~~&   --~~~~~&   --~~~~~&   --~~~~~&  52& PMNJ0408-7507        &   \\
 4.173&  76.908&   1.1 (0.2)&   --~~~~~&   --~~~~~&   --~~~~~&   --~~~~~&  53& NVSS J041045+765645  &   \\
 4.302&  38.044&   3.8 (0.4)&   3.4 (0.4)&   2.7 (0.3)&   3.3 (0.3)&   --~~~~~&    & GB6 J0418+3801       &   \\
 4.392&  -1.300&  10.6 (0.2)&   9.8 (0.3)&   9.4 (0.3)&   8.2 (0.3)&   4.0 (0.5)&  54& PMNJ0423-0120        &   \\
 4.399&  41.839&   1.4 (0.2)&   1.5 (0.3)&   --~~~~~&   --~~~~~&   --~~~~~&    & GB6 J0423+4150       &   \\
 4.412& -37.937&   1.5 (0.2)&   --~~~~~&   1.5 (0.2)&   1.3 (0.2)&   --~~~~~&  56& PMNJ0424-3756        &   \\
 4.416&   0.598&   1.9 (0.2)&   1.9 (0.2)&   1.8 (0.3)&   --~~~~~&   --~~~~~&  57& PMNJ0423+0031        &   \\
 4.479& -37.976&   1.2 (0.2)&   1.5 (0.2)&   1.2 (0.2)&   1.1 (0.2)&   --~~~~~&  58& PMNJ0428-3756        &   \\
 4.555&   5.377&   2.6 (0.2)&   2.8 (0.2)&   2.6 (0.2)&   2.2 (0.3)&   --~~~~~&  59& PMNJ0433+0521        &   \\
 4.615&  29.618&   3.5 (0.2)&   2.5 (0.2)&   1.8 (0.2)&   1.6 (0.3)&   --~~~~~&    & GB6 J0437+2940       &   \\
 4.633&  30.052&   --~~~~~&   --~~~~~&   1.3 (0.2)&   --~~~~~&   --~~~~~&    & GB6 J0438+3004       &   \\
 4.677& -43.529&   3.0 (0.2)&   2.5 (0.3)&   2.4 (0.2)&   1.7 (0.2)&   --~~~~~&  60& PMNJ0440-4332        &   \\
 4.709&  -0.250&   1.0 (0.2)&   --~~~~~&   1.3 (0.2)&   --~~~~~&   --~~~~~&  61& PMNJ0442-0017        &   \\
 4.820&  11.323&   2.2 (0.2)&   2.3 (0.2)&   2.3 (0.2)&   1.9 (0.3)&   --~~~~~&    & GB6 J0449+1121       &   \\
 4.832& -80.985&   2.0 (0.2)&   1.9 (0.2)&   1.5 (0.2)&   1.6 (0.2)&   --~~~~~&  62& PMNJ0450-8100        &   \\
 4.883& -28.159&   1.7 (0.2)&   1.6 (0.2)&   1.3 (0.2)&   1.2 (0.2)&   --~~~~~&  63& PMNJ0453-2807        &   \\
 4.925& -46.276&   3.9 (0.2)&   4.0 (0.2)&   4.1 (0.2)&   3.8 (0.2)&   2.1 (0.4)&  64& PMNJ0455-4616        & M \\
 4.947& -23.407&   2.8 (0.2)&   2.7 (0.2)&   2.7 (0.2)&   2.1 (0.2)&   --~~~~~&  65& PMNJ0457-2324        &   \\
 5.016&  -2.023&   1.3 (0.2)&   1.4 (0.2)&   1.5 (0.3)&   --~~~~~&   --~~~~~&  66& PMNJ0501-0159        &   \\
 5.114& -61.220&   2.3 (0.2)&   1.8 (0.2)&   1.6 (0.2)&   1.2 (0.2)&   --~~~~~&  67& PMNJ0506-6109        &   \\
 5.231& -22.010&   0.9 (0.2)&   --~~~~~&   --~~~~~&   --~~~~~&   --~~~~~&  69& PMNJ0513-2159        & M \\
 5.257& -45.940&   1.9 (0.2)&   --~~~~~&   --~~~~~&   --~~~~~&   --~~~~~&  70& PMNJ0515-4556        &   \\
 5.326& -45.736&   6.4 (0.3)&   5.0 (0.2)&   4.0 (0.2)&   2.8 (0.2)&   --~~~~~&  72& PMNJ0519-4546a       & M \\
 5.384& -36.484&   4.0 (0.2)&   3.4 (0.2)&   3.4 (0.2)&   3.1 (0.2)&   2.2 (0.4)&  73& PMNJ0522-3628        &   \\
 5.473&  21.500&   6.0 (0.8)&   4.3 (0.7)&   --~~~~~&   --~~~~~&   --~~~~~&    & NVSS J052830+213301  &   \\
 5.542&   7.487&   --~~~~~&   --~~~~~&   1.4 (0.3)&   --~~~~~&   --~~~~~&    & PMNJ0532+0732        &   \\
 5.554&  48.382&   1.2 (0.2)&   1.4 (0.2)&   --~~~~~&   --~~~~~&   --~~~~~&    & GB6 J0533+4822       &   \\
 5.651& -44.050&   5.6 (0.2)&   5.5 (0.2)&   5.4 (0.2)&   4.3 (0.2)&   2.5 (0.4)&  76& PMNJ0538-4405        &   \\
 5.710&  49.814&   1.7 (0.2)&   --~~~~~&   --~~~~~&   --~~~~~&   --~~~~~&  78& GB6 J0542+4951       &   \\
 5.929&  39.834&   2.8 (0.2)&   1.9 (0.2)&   --~~~~~&   --~~~~~&   --~~~~~&  80& GB6 J0555+3948       &   \\
 6.124&  67.357&   0.9 (0.2)&   --~~~~~&   --~~~~~&   --~~~~~&   --~~~~~&  82& GB6 J0607+6720       &   \\
 6.126&  -6.362&   7.0 (0.4)&   7.0 (0.6)&   8.9 (0.5)&   7.4 (0.4)&   7.8 (0.5)&    & PMNJ0607-0623        &   \\
 6.148& -22.383&   0.9 (0.2)&   --~~~~~&   --~~~~~&   --~~~~~&   --~~~~~&  83& PMNJ0608-2220        &   \\
 6.162& -15.727&   3.8 (0.2)&   3.5 (0.2)&   3.4 (0.2)&   2.3 (0.2)&   --~~~~~&  84& PMNJ0609-1542        &   \\
 6.451&  -5.874&   1.4 (0.2)&   --~~~~~&   --~~~~~&   --~~~~~&   --~~~~~&    & PMNJ0627-0553        &   \\
 6.486& -19.973&   1.4 (0.2)&   1.2 (0.2)&   1.6 (0.2)&   1.3 (0.2)&   --~~~~~&  86& PMNJ0629-1959        &   \\
 6.585& -75.258&   4.5 (0.3)&   3.8 (0.3)&   4.5 (0.2)&   3.2 (0.2)&   --~~~~~&  88& PMNJ0635-7516        &   \\
 6.608& -20.607&   1.2 (0.2)&   --~~~~~&   --~~~~~&   --~~~~~&   --~~~~~&  89& PMNJ0636-2041        & M \\
 6.653&  73.360&   --~~~~~&   --~~~~~&   1.1 (0.2)&   --~~~~~&   --~~~~~&  90& GB6 J0639+7324       &   \\
 6.776&  44.907&   3.2 (0.2)&   2.4 (0.2)&   2.1 (0.2)&   1.7 (0.3)&   --~~~~~&  91& GB6 J0646+4451       &   \\
 6.839& -16.564&   2.8 (0.2)&   2.4 (0.2)&   2.1 (0.2)&   1.6 (0.2)&   --~~~~~&    & PMNJ0650-1637        &   \\
 7.350&   4.029&   1.0 (0.2)&   --~~~~~&   --~~~~~&   --~~~~~&   --~~~~~&  92& PMNJ0721+0406        &   \\
 7.369&  71.363&   1.8 (0.2)&   1.8 (0.2)&   2.2 (0.2)&   1.9 (0.2)&   --~~~~~&  93& GB6 J0721+7120       &   \\
 7.418&  14.403&   1.0 (0.2)&   --~~~~~&   --~~~~~&   --~~~~~&   --~~~~~&    & GB6 J0725+1425       &   \\
 7.428&  -0.949&   1.2 (0.2)&   --~~~~~&   1.3 (0.2)&   --~~~~~&   --~~~~~&  94& PMNJ0725-0054        &   \\
 7.566&  50.305&   --~~~~~&   1.3 (0.3)&   --~~~~~&   --~~~~~&   --~~~~~&  96& GB6 J0733+5022       &   \\
 7.635&  17.668&   1.4 (0.2)&   1.4 (0.2)&   1.3 (0.2)&   --~~~~~&   --~~~~~&  97& GB6 J0738+1742       &   \\
 7.658&   1.627&   2.0 (0.2)&   2.2 (0.2)&   2.5 (0.2)&   2.3 (0.2)&   --~~~~~&  98& PMNJ0739+0137        &   \\
 7.684&  31.216&   1.3 (0.2)&   --~~~~~&   --~~~~~&   --~~~~~&   --~~~~~&  99& GB6 J0741+3112       &   \\
 7.719& -67.474&   1.4 (0.1)&   --~~~~~&   --~~~~~&   --~~~~~&   --~~~~~& 100& PMNJ0743-6726        &   \\
 7.763&  10.217&   1.3 (0.2)&   --~~~~~&   --~~~~~&   --~~~~~&   --~~~~~& 101& GB6 J0745+1011       &   \\
 7.768&  -0.713&   1.3 (0.2)&   --~~~~~&   --~~~~~&   --~~~~~&   --~~~~~& 102& PMNJ0745-0044        &   \\
 7.802& -36.071&   2.5 (0.4)&   1.9 (0.3)&   1.6 (0.2)&   --~~~~~&   --~~~~~&    & PMNJ0748-3605        &   \\
 7.847&  12.544&   2.7 (0.2)&   2.3 (0.3)&   2.6 (0.3)&   1.6 (0.3)&   --~~~~~& 103& GB6 J0750+1231       &   \\
 7.886&  53.824&   1.1 (0.2)&   --~~~~~&   --~~~~~&   --~~~~~&   --~~~~~& 104& GB6 J0753+5353       &   \\
 7.953&   9.971&   1.7 (0.2)&   1.7 (0.2)&   1.6 (0.3)&   --~~~~~&   --~~~~~& 105& PMNJ0757+0956        &   \\
 8.093&  -1.004&   1.0 (0.2)&   --~~~~~&   --~~~~~&   --~~~~~&   --~~~~~&    & NVSS J080537-005814  &   \\
 8.138&  -7.848&   1.4 (0.2)&   1.4 (0.2)&   1.4 (0.2)&   1.8 (0.3)&   --~~~~~& 106& PMNJ0808-0751        &   \\
 8.227&  48.188&   1.1 (0.2)&   --~~~~~&   --~~~~~&   --~~~~~&   --~~~~~& 107& GB6 J0813+4813       & M \\
 8.255&  36.557&   0.9 (0.2)&   --~~~~~&   --~~~~~&   --~~~~~&   --~~~~~&    & GB6 J0815+3635       &   \\
 8.411&  39.296&   1.3 (0.2)&   --~~~~~&   --~~~~~&   --~~~~~&   --~~~~~& 109& GB6 J0824+3916       &   \\
 8.428&   3.138&   1.4 (0.2)&   1.5 (0.2)&   1.4 (0.3)&   1.4 (0.3)&   --~~~~~& 110& PMNJ0825+0309        &   \\
 8.512&  24.154&   1.7 (0.2)&   1.4 (0.3)&   1.6 (0.3)&   2.0 (0.3)&   --~~~~~& 111& GB6 J0830+2410       &   \\
 8.614& -20.320&   2.9 (0.2)&   2.4 (0.2)&   2.1 (0.2)&   1.7 (0.3)&   --~~~~~& 112& PMNJ0836-2017        &   \\
 8.682&  13.226&   1.9 (0.2)&   2.1 (0.2)&   1.7 (0.3)&   --~~~~~&   --~~~~~& 114& GB6 J0840+1312       &   \\
 8.688&  70.908&   1.8 (0.2)&   1.8 (0.2)&   1.8 (0.2)&   1.6 (0.2)&   --~~~~~& 115& GB6 J0841+7053       &   \\
 8.917&  20.127&   3.9 (0.2)&   4.3 (0.2)&   3.9 (0.2)&   3.7 (0.3)&   --~~~~~& 116& GB6 J0854+2006       &   \\
 9.009& -28.100&   1.2 (0.2)&   --~~~~~&   --~~~~~&   --~~~~~&   --~~~~~&    & PMNJ0900-2808        &   \\
 9.041& -14.290&   1.2 (0.2)&   --~~~~~&   1.2 (0.2)&   --~~~~~&   --~~~~~& 117& PMNJ0902-1415        &   \\
 9.056&  46.835&   1.1 (0.2)&   --~~~~~&   --~~~~~&   --~~~~~&   --~~~~~&    & GB6 J0903+4650       &   \\
 9.078& -57.536&   1.1 (0.2)&   --~~~~~&   --~~~~~&   --~~~~~&   --~~~~~&    & PMNJ0904-5735        &   \\
 9.132& -20.302&   1.1 (0.2)&   --~~~~~&   --~~~~~&   --~~~~~&   --~~~~~& 118& PMNJ0906-2019        &   \\
 9.150&   1.320&   1.9 (0.2)&   1.9 (0.2)&   1.8 (0.3)&   1.5 (0.3)&   --~~~~~& 119& PMNJ0909+0121        &   \\
 9.157&  42.916&   --~~~~~&   --~~~~~&   1.3 (0.2)&   --~~~~~&   --~~~~~& 120& GB6 J0909+4253       &   \\
 9.299& -12.131&   2.2 (0.2)&   1.2 (0.2)&   1.1 (0.2)&   --~~~~~&   --~~~~~& 122& PMNJ0918-1205        &   \\
 9.357&  44.628&   1.3 (0.2)&   1.3 (0.2)&   1.2 (0.2)&   --~~~~~&   --~~~~~& 123& GB6 J0920+4441       & M \\
 9.362& -26.360&   1.5 (0.2)&   1.3 (0.2)&   1.1 (0.2)&   --~~~~~&   --~~~~~& 125& PMNJ0921-2618        &   \\
 9.365&  62.301&   0.9 (0.1)&   --~~~~~&   --~~~~~&   --~~~~~&   --~~~~~& 124& GB6 J0921+6215       &   \\
 9.378& -39.992&   1.3 (0.2)&   1.0 (0.2)&   1.0 (0.2)&   --~~~~~&   --~~~~~&    & PMNJ0922-3959        &   \\
 9.452&  39.010&   6.7 (0.2)&   5.3 (0.3)&   5.1 (0.2)&   4.0 (0.3)&   --~~~~~& 126& GB6 J0927+3902       &   \\
 9.816&  40.689&   1.5 (0.2)&   1.6 (0.2)&   1.2 (0.2)&   --~~~~~&   --~~~~~& 127& GB6 J0948+4039       &   \\
 9.939&  69.642&   1.4 (0.2)&   --~~~~~&   1.0 (0.2)&   --~~~~~&   --~~~~~& 128& GB6 J0955+6940       &   \\
 9.954&  55.444&   1.1 (0.2)&   --~~~~~&   --~~~~~&   --~~~~~&   --~~~~~& 129& GB6 J0957+5522       &   \\
 9.976&  47.405&   1.7 (0.2)&   1.4 (0.2)&   1.2 (0.2)&   --~~~~~&   --~~~~~& 130& GB6 J0958+4725       &   \\
 9.978& -41.221&   0.8 (0.2)&   --~~~~~&   --~~~~~&   --~~~~~&   --~~~~~&    & PMNJ0958-4110        &   \\
10.246& -45.101&   1.0 (0.2)&   --~~~~~&   --~~~~~&   --~~~~~&   --~~~~~&    & PMNJ1014-4508        &   \\
10.247&  23.040&   1.3 (0.2)&   --~~~~~&   --~~~~~&   --~~~~~&   --~~~~~& 131& GB6 J1014+2301       &   \\
10.410& -18.675&   --~~~~~&   --~~~~~&   1.1 (0.2)&   --~~~~~&   --~~~~~&    & PMNJ1024-1838        &   \\
10.620& -29.623&   1.4 (0.2)&   1.2 (0.2)&   1.1 (0.2)&   --~~~~~&   --~~~~~& 134& PMNJ1037-2934        &   \\
10.645&   5.191&   2.0 (0.2)&   2.0 (0.2)&   1.6 (0.2)&   1.5 (0.3)&   --~~~~~& 135& PMNJ1038+0512        &   \\
10.692& -47.705&   1.2 (0.2)&   --~~~~~&   --~~~~~&   --~~~~~&   --~~~~~& 137& PMNJ1041-4740        &   \\
10.692&   6.185&   1.5 (0.2)&   1.6 (0.2)&   1.5 (0.2)&   --~~~~~&   --~~~~~& 136& PMNJ1041+0610        &   \\
10.798&  71.789&   1.1 (0.2)&   --~~~~~&   --~~~~~&   --~~~~~&   --~~~~~& 138& GB6 J1048+7143       &   \\
10.804& -19.122&   1.2 (0.2)&   --~~~~~&   --~~~~~&   --~~~~~&   --~~~~~& 139& PMNJ1048-1909        &   \\
10.959& -80.042&   2.0 (0.2)&   2.1 (0.2)&   2.0 (0.2)&   2.1 (0.2)&   --~~~~~& 141& PMNJ1058-8003        &   \\
10.971&   1.533&   4.6 (0.2)&   4.3 (0.2)&   4.6 (0.2)&   4.0 (0.3)&   --~~~~~& 140& PMNJ1058+0133        &   \\
11.115& -44.857&   1.6 (0.2)&   1.5 (0.2)&   1.1 (0.2)&   1.3 (0.2)&   --~~~~~& 143& PMNJ1107-4449        & M \\
11.303& -46.591&   0.9 (0.2)&   --~~~~~&   --~~~~~&   --~~~~~&   --~~~~~& 144& PMNJ1118-4634        &   \\
11.314&  12.618&   1.1 (0.2)&   --~~~~~&   1.1 (0.2)&   --~~~~~&   --~~~~~& 145& GB6 J1118+1234       &   \\
11.448& -18.972&   1.6 (0.2)&   1.4 (0.2)&   1.3 (0.2)&   --~~~~~&   --~~~~~& 146& PMNJ1127-1857        &   \\
11.500& -14.735&   1.6 (0.2)&   1.6 (0.2)&   1.5 (0.2)&   --~~~~~&   --~~~~~& 147& PMNJ1130-1449        &   \\
11.514&  38.251&   1.5 (0.2)&   1.1 (0.2)&   --~~~~~&   --~~~~~&   --~~~~~& 148& GB6 J1130+3815       &   \\
11.756& -48.568&   0.7 (0.1)&   --~~~~~&   --~~~~~&   --~~~~~&   --~~~~~& 149& PMNJ1145-4836        &   \\
11.780&  39.953&   0.9 (0.2)&   --~~~~~&   --~~~~~&   --~~~~~&   --~~~~~& 150& GB6 J1146+3958       &   \\
11.785& -38.167&   1.9 (0.2)&   2.0 (0.2)&   1.9 (0.2)&   1.5 (0.3)&   --~~~~~& 151& PMNJ1147-3812        & M \\
11.890&  49.453&   2.0 (0.2)&   1.7 (0.2)&   1.9 (0.2)&   1.6 (0.2)&   --~~~~~& 153& GB6 J1153+4931       &   \\
11.913&  81.082&   1.2 (0.2)&   --~~~~~&   --~~~~~&   --~~~~~&   --~~~~~& 154& NVSS J115312+805829  & M \\
11.992&  29.230&   2.0 (0.2)&   2.2 (0.2)&   2.0 (0.2)&   1.8 (0.2)&   --~~~~~& 155& GB6 J1159+2914       &   \\
12.149& -24.126&   1.4 (0.2)&   --~~~~~&   --~~~~~&   --~~~~~&   --~~~~~& 157& PMNJ1209-2406        &   \\
12.179& -52.860&   2.6 (0.2)&   1.6 (0.2)&   1.1 (0.2)&   --~~~~~&   --~~~~~&    & PMNJ1211-5250        &   \\
12.262& -17.437&   1.7 (0.2)&   1.4 (0.2)&   --~~~~~&   --~~~~~&   --~~~~~& 158& PMNJ1215-1731        &   \\
12.322&   5.733&   2.4 (0.2)&   2.2 (0.3)&   2.0 (0.3)&   1.5 (0.3)&   --~~~~~& 160& PMNJ1219+0549        &   \\
12.484&   2.085&  18.7 (0.3)&  16.4 (0.3)&  15.2 (0.3)&  12.3 (0.3)&   5.5 (0.5)& 163& PMNJ1229+0203        &   \\
12.507&  12.055&  19.3 (0.3)&  14.7 (0.3)&  12.2 (0.2)&   8.1 (0.3)&   4.2 (0.4)& 164& GB6 J1230+1223       & M \\
12.762& -16.280&   0.9 (0.2)&   --~~~~~&   --~~~~~&   --~~~~~&   --~~~~~&    & PMNJ1245-1616        &   \\
12.781& -25.780&   1.4 (0.2)&   1.4 (0.2)&   1.8 (0.2)&   --~~~~~&   --~~~~~& 166& PMNJ1246-2547        &   \\
12.937&  -5.742&  16.7 (0.3)&  16.1 (0.3)&  17.4 (0.3)&  14.7 (0.3)&   7.9 (0.5)& 167& PMNJ1256-0547        &   \\
12.966& -31.981&   1.5 (0.1)&   1.1 (0.2)&   1.3 (0.2)&   --~~~~~&   --~~~~~& 168& PMNJ1257-3154        &   \\
12.979& -22.320&   1.0 (0.2)&   --~~~~~&   --~~~~~&   --~~~~~&   --~~~~~& 169& PMNJ1258-2219        &   \\
13.159&  11.942&   1.0 (0.2)&   --~~~~~&   --~~~~~&   --~~~~~&   --~~~~~& 172& GB6 J1309+1154       &   \\
13.176&  32.356&   3.2 (0.2)&   3.1 (0.2)&   2.8 (0.2)&   2.4 (0.2)&   --~~~~~& 173& GB6 J1310+3220       & M \\
13.266& -33.624&   1.6 (0.2)&   1.5 (0.2)&   1.6 (0.2)&   1.3 (0.2)&   --~~~~~& 174& PMNJ1316-3339        &   \\
13.351& -43.707&   --~~~~~&   2.4 (0.4)&   --~~~~~&   --~~~~~&   --~~~~~&    & PMNJ1321-4342        &   \\
13.388& -44.845&   3.5 (0.4)&   1.9 (0.4)&   --~~~~~&   --~~~~~&   --~~~~~&    & PMNJ1323-4452        &   \\
13.428& -43.042&  46.2 (0.5)&  36.7 (0.4)&  34.8 (0.3)&  25.2 (0.3)&  11.5 (0.4)&    & PMNJ1325-4257        & M \\
13.447& -52.981&   1.8 (0.2)&   1.8 (0.2)&   1.6 (0.2)&   1.8 (0.2)&   --~~~~~&    & PMNJ1326-5256        &   \\
13.453& -42.678&   --~~~~~&   --~~~~~&   7.3 (0.3)&   --~~~~~&   --~~~~~&    & PMNJ1327-4239        &   \\
13.457&  22.202&   1.1 (0.2)&   --~~~~~&   --~~~~~&   --~~~~~&   --~~~~~& 176& GB6 J1327+2210       &   \\
13.499&  31.859&   1.0 (0.2)&   --~~~~~&   --~~~~~&   --~~~~~&   --~~~~~& 177& GB6 J1329+3154       &   \\
13.516&  25.071&   1.1 (0.2)&   --~~~~~&   --~~~~~&   --~~~~~&   --~~~~~& 178& GB6 J1330+2509       &   \\
13.518&  30.570&   2.3 (0.2)&   1.9 (0.2)&   1.4 (0.2)&   --~~~~~&   --~~~~~& 179& GB6 J1331+3030       &   \\
13.552&   2.020&   1.3 (0.2)&   --~~~~~&   --~~~~~&   --~~~~~&   --~~~~~& 180& GB6 J1332+0200       &   \\
13.624& -34.130&   1.9 (0.2)&   1.2 (0.2)&   --~~~~~&   --~~~~~&   --~~~~~& 182& PMNJ1336-3358        & M \\
13.628& -12.897&   6.2 (0.2)&   6.2 (0.2)&   6.1 (0.2)&   5.3 (0.3)&   2.6 (0.4)& 183& PMNJ1337-1257        &   \\
13.743&  66.066&   0.9 (0.2)&   --~~~~~&   --~~~~~&   --~~~~~&   --~~~~~& 184& GB6 J1344+6606       &   \\
13.794&  12.303&   0.9 (0.2)&   1.1 (0.2)&   --~~~~~&   --~~~~~&   --~~~~~&    & GB6 J1347+1217       &   \\
13.912& -10.653&   1.7 (0.2)&   1.3 (0.2)&   1.7 (0.2)&   --~~~~~&   --~~~~~& 185& PMNJ1354-1041        &   \\
13.951&  76.746&   0.8 (0.2)&   --~~~~~&   --~~~~~&   --~~~~~&   --~~~~~& 187& NVSS J135755+764320  &   \\
13.952&  19.315&   1.4 (0.2)&   1.4 (0.2)&   1.1 (0.2)&   --~~~~~&   --~~~~~& 186& GB6 J1357+1919       &   \\
13.952& -15.488&   1.1 (0.2)&   --~~~~~&   --~~~~~&   --~~~~~&   --~~~~~& 188& PMNJ1357-1527        &   \\
14.148&  -7.826&   1.3 (0.2)&   1.1 (0.2)&   1.2 (0.2)&   --~~~~~&   --~~~~~& 189& NVSS J140856-075226  &   \\
14.268&  13.325&   0.9 (0.2)&   --~~~~~&   --~~~~~&   --~~~~~&   --~~~~~&    & GB6 J1415+1320       &   \\
14.327&  38.385&   0.9 (0.2)&   1.0 (0.2)&   --~~~~~&   1.0 (0.2)&   --~~~~~& 192& GB6 J1419+3822       & M \\
14.327&  54.355&   1.0 (0.2)&   --~~~~~&   --~~~~~&   1.2 (0.2)&   --~~~~~& 191& GB6 J1419+5423       &   \\
14.334&  27.040&   1.1 (0.1)&   --~~~~~&   --~~~~~&   --~~~~~&   --~~~~~& 193& NVSS J141958+270143  &   \\
14.408& -49.221&   2.2 (0.2)&   1.6 (0.2)&   1.5 (0.2)&   --~~~~~&   --~~~~~&    & PMNJ1424-4913        &   \\
14.412& -68.110&   1.1 (0.2)&   --~~~~~&   1.2 (0.2)&   --~~~~~&   --~~~~~&    & PMNJ1424-6808        &   \\
14.457& -33.351&   1.0 (0.2)&   1.5 (0.2)&   1.8 (0.2)&   1.5 (0.3)&   --~~~~~& 194& NVSS J142717-331800  &   \\
14.463& -42.116&   3.2 (0.2)&   2.9 (0.2)&   2.8 (0.2)&   2.3 (0.3)&   --~~~~~& 195& PMNJ1427-4206        &   \\
14.635& -22.120&   0.8 (0.2)&   1.1 (0.2)&   --~~~~~&   --~~~~~&   --~~~~~&    & PMNJ1438-2204        &   \\
14.714&  52.080&   1.0 (0.2)&   --~~~~~&   --~~~~~&   --~~~~~&   --~~~~~& 197& GB6 J1443+5201       &   \\
14.909& -37.741&   1.0 (0.2)&   --~~~~~&   --~~~~~&   --~~~~~&   --~~~~~&    & PMNJ1454-3747        &   \\
14.991&  71.762&   1.5 (0.2)&   1.6 (0.2)&   1.0 (0.2)&   --~~~~~&   --~~~~~& 198& GB6 J1459+7140       &   \\
15.064& -41.927&   2.3 (0.2)&   1.6 (0.2)&   1.6 (0.2)&   --~~~~~&   --~~~~~&    & PMNJ1503-4154        &   \\
15.078&  10.490&   2.0 (0.2)&   1.7 (0.2)&   1.5 (0.2)&   --~~~~~&   --~~~~~& 199& GB6 J1504+1029       &   \\
15.119& -16.887&   1.6 (0.2)&   1.5 (0.2)&   --~~~~~&   --~~~~~&   --~~~~~& 200& PMNJ1507-1652        &   \\
15.180&  -5.634&   1.1 (0.2)&   --~~~~~&   --~~~~~&   --~~~~~&   --~~~~~& 201& PMNJ1510-0543        &   \\
15.211&  -9.147&   1.8 (0.2)&   1.7 (0.2)&   1.8 (0.2)&   1.7 (0.3)&   --~~~~~& 202& NVSS J151250-090600  &   \\
15.228& -10.228&   1.0 (0.2)&   --~~~~~&   --~~~~~&   --~~~~~&   --~~~~~& 203& PMNJ1513-1012        &   \\
15.248& -47.857&   1.7 (0.2)&   1.4 (0.2)&   1.4 (0.2)&   --~~~~~&   --~~~~~&    & PMNJ1514-4748        &   \\
15.285&   0.206&   1.7 (0.2)&   1.8 (0.2)&   1.6 (0.2)&   --~~~~~&   --~~~~~& 204& PMNJ1516+0014        &   \\
15.296& -24.422&   2.0 (0.2)&   1.9 (0.2)&   1.9 (0.2)&   1.6 (0.3)&   --~~~~~& 205& PMNJ1517-2422        &   \\
15.380& -27.510&   1.3 (0.3)&   --~~~~~&   1.2 (0.2)&   --~~~~~&   --~~~~~&    & PMNJ1522-2730        &   \\
15.680&  14.846&   1.2 (0.2)&   --~~~~~&   --~~~~~&   --~~~~~&   --~~~~~& 206& GB6 J1540+1447       &   \\
15.825&  50.657&   1.1 (0.2)&   --~~~~~&   0.9 (0.2)&   --~~~~~&   --~~~~~& 207& GB6 J1549+5038       &   \\
15.826&   2.660&   2.3 (0.2)&   2.3 (0.3)&   2.0 (0.2)&   1.8 (0.3)&   --~~~~~& 208& PMNJ1549+0237        &   \\
15.845&   5.469&   2.6 (0.2)&   2.2 (0.3)&   1.8 (0.3)&   1.9 (0.3)&   --~~~~~& 209& PMNJ1550+0527        & M \\
16.042&  33.446&   0.9 (0.1)&   --~~~~~&   --~~~~~&   --~~~~~&   --~~~~~& 211& GB6 J1602+3326       & M \\
16.150&  10.456&   2.6 (0.2)&   2.4 (0.2)&   2.0 (0.2)&   1.7 (0.3)&   --~~~~~& 212& GB6 J1608+1029       &   \\
16.228&  34.206&   3.9 (0.2)&   3.3 (0.2)&   3.2 (0.2)&   2.7 (0.2)&   --~~~~~& 213& GB6 J1613+3412       &   \\
16.250& -60.657&   2.6 (0.2)&   1.8 (0.3)&   --~~~~~&   --~~~~~&   --~~~~~&    & PMNJ1615-6054        &   \\
16.286& -58.815&   2.6 (0.4)&   --~~~~~&   2.1 (0.4)&   --~~~~~&   --~~~~~&    & PMNJ1617-5848        &   \\
16.310& -77.341&   2.4 (0.2)&   2.2 (0.2)&   1.9 (0.2)&   1.7 (0.2)&   --~~~~~& 214& PMNJ1617-7717        & M \\
16.431& -25.425&   --~~~~~&   --~~~~~&   2.3 (0.3)&   2.1 (0.3)&   --~~~~~&    & PMNJ1625-2527        &   \\
16.431& -29.903&   1.6 (0.3)&   --~~~~~&   --~~~~~&   --~~~~~&   --~~~~~&    & NVSS J162605-295126  &   \\
16.539&  82.520&   1.2 (0.1)&   1.3 (0.2)&   1.2 (0.2)&   --~~~~~&   --~~~~~& 215& NVSS J163051+823345  &   \\
16.581&  38.150&   4.2 (0.2)&   4.7 (0.2)&   4.5 (0.2)&   3.6 (0.2)&   2.2 (0.4)& 216& GB6 J1635+3808       &   \\
16.636&  47.260&   --~~~~~&   --~~~~~&   1.0 (0.2)&   --~~~~~&   --~~~~~& 217& GB6 J1637+4717       &   \\
16.640&  57.362&   1.3 (0.2)&   1.4 (0.2)&   1.5 (0.2)&   1.4 (0.2)&   --~~~~~& 218& GB6 J1638+5720       &   \\
16.676&  39.774&   --~~~~~&   --~~~~~&   1.8 (0.2)&   --~~~~~&   --~~~~~&    & GB6 J1640+3947       &   \\
16.708&  68.891&   1.2 (0.2)&   1.0 (0.2)&   1.2 (0.2)&   1.3 (0.2)&   --~~~~~& 219& GB6 J1642+6856       &   \\
16.719&  39.812&   7.0 (0.3)&   6.2 (0.2)&   6.0 (0.2)&   5.2 (0.3)&   2.7 (0.4)& 220& GB6 J1642+3948       &   \\
16.855&   4.951&   1.9 (0.2)&   1.3 (0.2)&   1.2 (0.2)&   --~~~~~&   --~~~~~& 223& PMNJ1651+0459        &   \\
16.897&  39.715&   1.2 (0.2)&   --~~~~~&   --~~~~~&   --~~~~~&   --~~~~~& 224& GB6 J1653+3945       & M \\
16.965&  47.823&   0.9 (0.2)&   --~~~~~&   --~~~~~&   --~~~~~&   --~~~~~& 226& GB6 J1658+4737       &   \\
16.969&   7.628&   1.1 (0.2)&   --~~~~~&   --~~~~~&   --~~~~~&   --~~~~~& 227& PMNJ1658+0741        &   \\
17.026& -56.367&   1.2 (0.2)&   --~~~~~&   --~~~~~&   --~~~~~&   --~~~~~&    & PMNJ1701-5621        &   \\
17.064& -62.252&   1.8 (0.2)&   1.8 (0.2)&   1.9 (0.2)&   1.4 (0.2)&   --~~~~~& 229& PMNJ1703-6212        & M \\
17.341&  -1.005&   5.7 (0.2)&   3.9 (0.2)&   3.3 (0.2)&   2.6 (0.3)&   --~~~~~&    & PMNJ1720-0058        & M \\
17.388& -65.003&   2.4 (0.2)&   1.9 (0.2)&   1.5 (0.2)&   1.2 (0.2)&   --~~~~~& 232& PMNJ1723-6500        &   \\
17.453&  45.540&   1.1 (0.2)&   1.3 (0.2)&   --~~~~~&   1.3 (0.2)&   --~~~~~& 233& GB6 J1727+4530       & M \\
17.488& -23.821&   --~~~~~&   --~~~~~&   2.1 (0.4)&   --~~~~~&   --~~~~~&    & PMNJ1729-2345        &   \\
17.515& -21.458&   2.2 (0.4)&   1.7 (0.3)&   1.9 (0.4)&   --~~~~~&   --~~~~~&    & PMNJ1730-2129        &   \\
17.551& -13.075&   5.4 (0.2)&   5.4 (0.2)&   5.1 (0.2)&   4.0 (0.3)&   --~~~~~&    & PMNJ1733-1304        &   \\
17.570&  39.009&   1.1 (0.2)&   1.3 (0.2)&   1.1 (0.2)&   1.2 (0.2)&   --~~~~~& 234& GB6 J1734+3857       &   \\
17.625&   6.316&   1.3 (0.2)&   --~~~~~&   1.1 (0.2)&   --~~~~~&   --~~~~~&    & GB6 J1737+0620       &   \\
17.663& -71.251&   --~~~~~&   --~~~~~&   0.9 (0.2)&   --~~~~~&   --~~~~~&    & SUMSS J173939-712325 &   \\
17.667&  47.672&   --~~~~~&   --~~~~~&   1.0 (0.2)&   --~~~~~&   --~~~~~& 236& GB6 J1739+4738       &   \\
17.678&  52.175&   1.3 (0.2)&   1.1 (0.2)&   1.3 (0.2)&   1.4 (0.2)&   --~~~~~& 237& GB6 J1740+5211      5&   \\
17.731&  -3.789&   5.8 (0.2)&   5.2 (0.2)&   4.9 (0.2)&   4.5 (0.3)&   --~~~~~&    & PMNJ1743-0350        &   \\
17.739& -51.762&   1.0 (0.2)&   --~~~~~&   --~~~~~&   --~~~~~&   --~~~~~&    & PMNJ1744-5144        &   \\
17.797&  70.113&   --~~~~~&   --~~~~~&   --~~~~~&   1.0 (0.2)&   --~~~~~& 238& GB6 J1748+7005       &   \\
17.860&   9.709&   4.1 (0.2)&   4.2 (0.2)&   4.3 (0.2)&   4.2 (0.2)&   2.5 (0.4)&    & GB6 J1751+0938       & M \\
17.879&   9.941&   --~~~~~&   --~~~~~&   --~~~~~&   2.0 (0.2)&   --~~~~~&    & NVSS J175231+095711  &   \\
17.884&   9.351&   --~~~~~&   --~~~~~&   1.4 (0.2)&   --~~~~~&   --~~~~~&    & NVSS J175302+091959  &   \\
17.888&  44.131&   0.9 (0.2)&   --~~~~~&   --~~~~~&   --~~~~~&   --~~~~~& 239& GB6 J1753+4410       & M \\
17.893&  28.730&   2.3 (0.1)&   2.2 (0.2)&   2.3 (0.2)&   2.3 (0.2)&   --~~~~~& 240& GB6 J1753+2847       &   \\
18.008&  38.842&   1.0 (0.1)&   --~~~~~&   --~~~~~&   --~~~~~&   --~~~~~& 242& GB6 J1800+3848       & M \\
18.014&  78.470&   2.1 (0.1)&   1.6 (0.2)&   1.7 (0.2)&   1.5 (0.2)&   --~~~~~& 243& NVSS J180045+782805  &   \\
18.023&  44.067&   1.1 (0.2)&   1.1 (0.2)&   1.4 (0.2)&   1.4 (0.2)&   --~~~~~& 244& GB6 J1801+4404       &   \\
18.047& -39.695&   1.9 (0.2)&   1.9 (0.2)&   1.6 (0.2)&   2.0 (0.3)&   --~~~~~&    & PMNJ1802-3940        &   \\
18.056& -65.156&   1.3 (0.2)&   1.2 (0.2)&   1.5 (0.2)&   --~~~~~&   --~~~~~& 245& PMNJ1803-6507        & M \\
18.119&  69.806&   1.6 (0.1)&   1.5 (0.2)&   1.3 (0.2)&   1.5 (0.2)&   --~~~~~& 246& GB6 J1806+6949       &   \\
18.166& -45.918&   1.5 (0.2)&   1.5 (0.2)&   1.3 (0.2)&   1.3 (0.3)&   --~~~~~&    & PMNJ1809-4552        &   \\
18.205&   6.838&   1.1 (0.2)&   --~~~~~&   --~~~~~&   --~~~~~&   --~~~~~&    & GB6 J1812+0651       &   \\
18.324& -55.350&   0.9 (0.2)&   --~~~~~&   --~~~~~&   --~~~~~&   --~~~~~& 247& PMNJ1819-5521        &   \\
18.363& -63.968&   1.6 (0.2)&   1.5 (0.2)&   --~~~~~&   --~~~~~&   --~~~~~& 248& PMNJ1819-6345        &   \\
18.398&  56.812&   1.5 (0.2)&   1.4 (0.2)&   1.4 (0.2)&   1.2 (0.2)&   --~~~~~& 249& GB6 J1824+5650       &   \\
18.495&  48.771&   2.7 (0.2)&   2.4 (0.2)&   2.4 (0.2)&   1.6 (0.2)&   --~~~~~& 250& GB6 J1829+4844       &   \\
18.562& -21.110&   4.9 (0.5)&   4.5 (0.5)&   4.6 (0.5)&   2.7 (0.4)&   --~~~~~&    & PMNJ1833-2103        &   \\
18.584&  32.650&   0.9 (0.1)&   --~~~~~&   --~~~~~&   --~~~~~&   --~~~~~& 251& GB6 J1835+3241       &   \\
18.609& -71.853&   0.8 (0.1)&   --~~~~~&   --~~~~~&   --~~~~~&   --~~~~~&    & PMNJ1835-7150        &   \\
18.626& -71.134&   1.8 (0.1)&   1.5 (0.2)&   1.3 (0.2)&   1.0 (0.2)&   --~~~~~& 252& PMNJ1837-7108        &   \\
18.719&  79.780&   1.1 (0.2)&   --~~~~~&   --~~~~~&   --~~~~~&   --~~~~~& 253& NVSS J184226+794517  &   \\
18.722&  68.146&   1.0 (0.2)&   1.2 (0.2)&   1.1 (0.2)&   --~~~~~&   --~~~~~& 254& GB6 J1842+6809       &   \\
18.824&  67.155&   1.4 (0.2)&   1.5 (0.2)&   1.3 (0.2)&   1.0 (0.2)&   --~~~~~& 256& GB6 J1849+6705       &   \\
18.842&  28.396&   1.2 (0.2)&   --~~~~~&   --~~~~~&   --~~~~~&   --~~~~~& 257& GB6 J1850+2825       &   \\
19.051&  31.972&   1.0 (0.2)&   --~~~~~&   --~~~~~&   --~~~~~&   --~~~~~& 258& GB6 J1902+3159       &   \\
19.183& -20.091&   2.0 (0.2)&   1.9 (0.3)&   1.7 (0.3)&   1.7 (0.3)&   --~~~~~&    & PMNJ1911-2006        &   \\
19.394& -21.033&   2.3 (0.3)&   2.1 (0.3)&   2.0 (0.3)&   2.0 (0.3)&   --~~~~~& 259& PMNJ1923-2104        &   \\
19.417& -29.201&  12.1 (0.2)&  10.9 (0.3)&  10.6 (0.2)&   9.6 (0.3)&   4.4 (0.4)&    & PMNJ1924-2914        & M \\
19.455&  61.295&   1.2 (0.2)&   1.1 (0.2)&   1.2 (0.2)&   --~~~~~&   --~~~~~& 260& GB6 J1927+6117       &   \\
19.460&  73.985&   3.6 (0.2)&   3.2 (0.2)&   2.6 (0.2)&   2.5 (0.2)&   --~~~~~& 261& GB6 J1927+7357       &   \\
19.621& -40.005&   --~~~~~&   --~~~~~&   1.2 (0.2)&   --~~~~~&   --~~~~~& 262& PMNJ1937-3957        &   \\
19.657& -15.422&   1.1 (0.2)&   --~~~~~&   --~~~~~&   --~~~~~&   --~~~~~&    & PMNJ1939-1525        & M \\
19.659& -63.710&   0.9 (0.2)&   --~~~~~&   --~~~~~&   --~~~~~&   --~~~~~& 263& PMNJ1939-6342        &   \\
19.659& -69.144&   0.7 (0.1)&   --~~~~~&   --~~~~~&   --~~~~~&   --~~~~~&    & PMNJ1940-6908        &   \\
19.872&   2.514&   1.1 (0.2)&   --~~~~~&   --~~~~~&   --~~~~~&   --~~~~~& 264& PMNJ1952+0230        &   \\
19.934&  51.548&   --~~~~~&   1.1 (0.2)&   --~~~~~&   --~~~~~&   --~~~~~& 265& GB6 J1955+5131       &   \\
19.961& -38.736&   3.3 (0.2)&   3.2 (0.2)&   2.9 (0.2)&   2.5 (0.3)&   --~~~~~& 266& PMNJ1957-3845        &   \\
19.964& -55.167&   0.9 (0.2)&   --~~~~~&   --~~~~~&   --~~~~~&   --~~~~~&    & PMNJ1958-5509        &   \\
19.997&  40.704&  54.6 (0.4)& 34.9 (0.4)&  30.4 (0.4)&  18.0 (0.4)&  7.9 (0.5)&  &  NVSS J195921+403428  &  M \\
20.018& -17.850&   1.8 (0.2)&   1.5 (0.2)&   1.3 (0.2)&   1.5 (0.3)&   --~~~~~& 267& PMNJ2000-1748        &   \\
20.100&  77.924&   --~~~~~&   1.0 (0.2)&   0.9 (0.2)&   --~~~~~&   --~~~~~& 268& NVSS J200531+775243  &   \\
20.159& -48.794&   0.8 (0.2)&   --~~~~~&   --~~~~~&   --~~~~~&   --~~~~~&    & PMNJ2009-4849        &   \\
20.190& -15.743&   1.6 (0.2)&   1.5 (0.3)&   1.5 (0.2)&   --~~~~~&   --~~~~~& 270& PMNJ2011-1546        &   \\
20.372&  61.679&   1.6 (0.2)&   1.4 (0.2)&   1.1 (0.2)&   --~~~~~&   --~~~~~& 271& GB6 J2022+6137       &   \\
20.415&  17.256&   0.9 (0.2)&   --~~~~~&   --~~~~~&   --~~~~~&   --~~~~~&    & GB6 J2024+1718       &   \\
20.609& -68.771&   0.9 (0.1)&   1.1 (0.2)&   --~~~~~&   --~~~~~&   --~~~~~& 272& PMNJ2035-6846        &   \\
20.640&  51.296&   2.4 (0.3)&   --~~~~~&   1.5 (0.2)&   1.7 (0.3)&   --~~~~~&    & GB6 J2038+5119       &   \\
20.933& -47.226&   1.6 (0.2)&   1.5 (0.2)&   1.7 (0.2)&   1.5 (0.3)&   --~~~~~& 274& PMNJ2056-4714        &   \\
21.024&   3.682&   0.8 (0.2)&   --~~~~~&   --~~~~~&   --~~~~~&   --~~~~~&    & PMNJ2101+0341        &   \\
21.034& -27.992&   1.0 (0.2)&   --~~~~~&   --~~~~~&   --~~~~~&   --~~~~~&    & PMNJ2101-2802        &   \\
21.159&  35.494&   1.2 (0.2)&   --~~~~~&   --~~~~~&   --~~~~~&   --~~~~~& 276& GB6 J2109+3532       &   \\
21.166& -41.208&   1.3 (0.2)&   1.6 (0.2)&   1.2 (0.2)&   --~~~~~&   --~~~~~& 275& PMNJ2109-4110        &   \\
21.394&   5.611&   2.2 (0.2)&   1.8 (0.2)&   1.7 (0.2)&   1.3 (0.3)&   --~~~~~& 277& PMNJ2123+0535        &   \\
21.396&  25.014&   0.9 (0.2)&   --~~~~~&   --~~~~~&   --~~~~~&   --~~~~~&    & GB6 J2123+2504       &   \\
21.529& -12.153&   2.9 (0.2)&   2.3 (0.2)&   2.1 (0.2)&   1.8 (0.3)&   --~~~~~& 278& PMNJ2131-1207        &   \\
21.570&  -1.904&   1.9 (0.2)&   1.8 (0.2)&   1.7 (0.3)&   1.5 (0.3)&   --~~~~~& 279& PMNJ2134-0153        &   \\
21.606&   0.734&   4.5 (0.3)&   3.7 (0.3)&   2.9 (0.3)&   --~~~~~&   --~~~~~& 280& PMNJ2136+0041        &   \\
21.651&  14.410&   2.2 (0.2)&   1.9 (0.2)&   1.6 (0.2)&   --~~~~~&   --~~~~~& 281& GB6 J2139+1423       &   \\
21.730&  17.741&   1.2 (0.2)&   1.1 (0.2)&   --~~~~~&   --~~~~~&   --~~~~~& 282& GB6 J2143+1743       & M \\
21.762& -77.935&   1.3 (0.2)&   1.1 (0.2)&   --~~~~~&   --~~~~~&   --~~~~~& 284& PMNJ2146-7755        &   \\
21.803&   7.015&   8.1 (0.2)&   7.2 (0.2)&   6.9 (0.2)&   5.6 (0.3)&   3.0 (0.4)& 283& PMNJ2148+0657        &   \\
21.855&   7.113&   1.3 (0.2)&   --~~~~~&   --~~~~~&   --~~~~~&   --~~~~~&    & PMNJ2151+0709        &   \\
21.872& -30.498&   1.1 (0.2)&   1.3 (0.2)&   1.2 (0.2)&   1.5 (0.3)&   --~~~~~& 285& PMNJ2151-3028        &   \\
21.888&  47.306&   2.5 (0.2)&   2.7 (0.3)&   2.2 (0.2)&   2.1 (0.2)&   2.0 (0.4)&    & GB6 J2153+4716       &   \\
21.962& -69.639&   3.8 (0.2)&   2.8 (0.2)&   2.6 (0.2)&   1.9 (0.2)&   --~~~~~& 286& PMNJ2157-6941        &   \\
21.969& -15.057&   2.3 (0.2)&   2.0 (0.2)&   1.8 (0.2)&   --~~~~~&   --~~~~~& 287& PMNJ2158-1501        &   \\
22.048&  42.317&   3.1 (0.2)&   2.8 (0.2)&   2.9 (0.2)&   2.4 (0.2)&   --~~~~~& 288& GB6 J2202+4216       &   \\
22.053&  31.777&   2.9 (0.2)&   2.8 (0.2)&   2.1 (0.2)&   1.7 (0.2)&   --~~~~~& 289& GB6 J2203+3145       &   \\
22.059&  17.410&   1.7 (0.2)&   1.6 (0.2)&   1.6 (0.2)&   1.7 (0.3)&   --~~~~~& 290& GB6 J2203+1725       &   \\
22.101& -18.642&   1.8 (0.2)&   1.5 (0.2)&   1.2 (0.2)&   --~~~~~&   --~~~~~& 291& PMNJ2206-1835        &   \\
22.133& -53.712&   1.1 (0.2)&   --~~~~~&   --~~~~~&   --~~~~~&   --~~~~~& 292& PMNJ2207-5346        &   \\
22.199&  23.874&   1.1 (0.2)&   1.3 (0.2)&   --~~~~~&   --~~~~~&   --~~~~~& 293& GB6 J2212+2355       &   \\
22.218& -25.529&   0.9 (0.2)&   --~~~~~&   --~~~~~&   --~~~~~&   --~~~~~&    & PMNJ2213-2529        & M \\
22.314&  -3.630&   2.7 (0.2)&   2.0 (0.2)&   2.0 (0.4)&   --~~~~~&   --~~~~~& 294& PMNJ2218-0335        &   \\
22.426&  21.289&   --~~~~~&   1.2 (0.2)&   --~~~~~&   --~~~~~&   --~~~~~&    & GB6 J2225+2118       &   \\
22.427&  -4.985&   5.0 (0.2)&   4.6 (0.3)&   3.9 (0.3)&   3.5 (0.3)&   --~~~~~& 295& PMNJ2225-0457        &   \\
22.498&  -8.534&   2.5 (0.2)&   2.6 (0.3)&   2.7 (0.3)&   3.0 (0.3)&   --~~~~~& 296& PMNJ2229-0832        & M \\
22.547&  11.767&   3.4 (0.2)&   3.2 (0.2)&   3.4 (0.2)&   3.1 (0.3)&   2.6 (0.5)& 298& GB6 J2232+1143       &   \\
22.589& -48.650&   1.9 (0.2)&   1.8 (0.2)&   1.7 (0.2)&   1.5 (0.2)&   --~~~~~& 299& PMNJ2235-4835        &   \\
22.603&  28.502&   1.3 (0.2)&   1.3 (0.2)&   --~~~~~&   --~~~~~&   --~~~~~& 300& GB6 J2236+2828       &   \\
22.648& -56.994&   0.9 (0.2)&   --~~~~~&   --~~~~~&   --~~~~~&   --~~~~~& 301& PMNJ2239-5701        &   \\
22.774& -12.133&   1.2 (0.2)&   1.2 (0.2)&   1.1 (0.2)&   --~~~~~&   --~~~~~& 302& PMNJ2246-1206        &   \\
22.901&  16.135&   7.4 (0.2)&   6.7 (0.2)&   6.8 (0.2)&   6.1 (0.3)&   4.0 (0.4)& 303& GB6 J2253+1608       &   \\
22.958&  41.923&   1.0 (0.2)&   --~~~~~&   --~~~~~&   --~~~~~&   --~~~~~&    & GB6 J2257+4154       &   \\
22.970& -28.003&   6.8 (0.2)&   6.4 (0.2)&   6.2 (0.2)&   5.6 (0.3)&   3.5 (0.4)& 306& PMNJ2258-2758        &   \\
23.367&  27.500&   0.9 (0.2)&   --~~~~~&   --~~~~~&   --~~~~~&   --~~~~~&    & GB6 J2322+2732       &   \\
23.395&  -3.296&   1.0 (0.2)&   --~~~~~&   --~~~~~&   --~~~~~&   --~~~~~&    & PMNJ2323-0317        &   \\
23.457&   9.663&   --~~~~~&   1.4 (0.2)&   1.3 (0.2)&   --~~~~~&   --~~~~~& 309& PMNJ2327+0940        &   \\
23.489& -47.438&   1.3 (0.2)&   1.1 (0.2)&   1.3 (0.2)&   --~~~~~&   --~~~~~& 310& PMNJ2329-4730        &   \\
23.512&  10.973&   1.0 (0.2)&   --~~~~~&   --~~~~~&   --~~~~~&   --~~~~~& 311& GB6 J2330+1100       &   \\
23.528& -15.973&   1.4 (0.2)&   --~~~~~&   --~~~~~&   --~~~~~&   --~~~~~& 312& PMNJ2331-1556        & M \\
23.566& -23.721&   --~~~~~&   --~~~~~&   1.2 (0.2)&   --~~~~~&   --~~~~~& 313& PMNJ2333-2343        &   \\
23.568&   7.559&   1.2 (0.2)&   --~~~~~&   --~~~~~&   --~~~~~&   --~~~~~& 314& PMNJ2334+0736        &   \\
23.589&  -1.493&   --~~~~~&   1.2 (0.2)&   1.3 (0.2)&   --~~~~~&   --~~~~~& 315& PMNJ2335-0131        &   \\
23.602& -52.602&   0.9 (0.2)&   --~~~~~&   --~~~~~&   --~~~~~&   --~~~~~& 316& PMNJ2336-5236        &   \\
23.778&   9.453&   1.1 (0.2)&   --~~~~~&   --~~~~~&   --~~~~~&   --~~~~~& 317& PMNJ2346+0930        &   \\
23.803& -16.538&   1.9 (0.2)&   1.7 (0.3)&   2.0 (0.2)&   1.6 (0.3)&   --~~~~~& 318& PMNJ2348-1631        &   \\
23.907&  45.943&   1.7 (0.1)&   1.2 (0.2)&   1.2 (0.2)&   --~~~~~&   --~~~~~& 319& GB6 J2354+4553       &   \\
23.964& -53.162&   1.3 (0.1)&   1.1 (0.2)&   1.2 (0.2)&   --~~~~~&   --~~~~~& 321& PMNJ2357-5311        &   \\
23.964& -10.295&   1.0 (0.2)&   --~~~~~&   --~~~~~&   --~~~~~&   --~~~~~& 322& PMNJ2358-1020        &   \\
23.979& -60.967&   1.8 (0.2)&   1.4 (0.2)&   1.1 (0.2)&   --~~~~~&   --~~~~~& 323& PMNJ2358-6054        &   \\
\enddata
\end{deluxetable*}

\end{document}